\documentclass[aps, prl, amsfonts, amsmath, amssymb, reprint, showkeys, nofootinbib, twoside, superscriptaddress, letterpaper]{revtex4-2}

\usepackage{graphicx}% Include figure files
\usepackage{dcolumn}% Align table columns on decimal point
\usepackage{bm}% bold math
\usepackage{hyperref}% add hypertext capabilities
\usepackage{soul}
\hypersetup{
    colorlinks=true,       % false: boxed links; true: colored text links
    linkcolor=blue,        % color of internal links (e.g., \ref, \pageref)
    citecolor=blue,        % color of citation links (e.g., \cite)
    filecolor=magenta,     % color of file links
    urlcolor=blue          % color of external hyperlinks
}
\usepackage{lipsum}
\usepackage[version=4]{mhchem}
\usepackage{amsmath,amssymb,amsfonts}%
\usepackage{siunitx}%
\DeclareSIUnit{\wv}{\%\,w/v}
\DeclareSIUnit{\square}{sq}
\DeclareSIUnit{\pix}{pixel}
\DeclareSIUnit{\fps}{fps}
\DeclareSIUnit{\radians}{rad}
\DeclareSIUnit{\elecVolt}{eV}
\DeclareSIUnit{\molar}{M}
\usepackage[dvipsnames]{xcolor}

\usepackage{comment}

\usepackage{utfsym}
\newcommand{\newcheckmark}{\usym{2713}}
\newcommand{\newcrossmark}{\usym{2717}}

\begin{document}

\preprint{}
\title{Universal transport of active colloids with sensory delay in motility landscapes}

\author{Adri\`a Garc\'es}
    \email{adria.garces@ub.edu}
    \affiliation{Departament de Física de la Matèria Condensada, Universitat de Barcelona, Mart\'i i Franqu\`es 1, E08028 Barcelona, Spain}
    \affiliation{University of Barcelona Institute of Complex Systems (UBICS), Mart\'i i Franqu\`es 1, E08028 Barcelona, Spain}
    \affiliation{Computing and Understanding Collective Action (CUCA) Lab, Mart\'i i Franqu\`es 1, E08028 Barcelona, Spain}

\author{Ueli T\"opfer}
    \email{ueli.toepfer@mat.ethz.ch}
    \affiliation{Department of Materials, ETH Zurich, 8049 Zurich, Switzerland}
    
\author{Lucio Isa}
    \affiliation{Department of Materials, ETH Zurich, 8049 Zurich, Switzerland}

\author{Demian Levis}
    \affiliation{Departament de Física de la Matèria Condensada, Universitat de Barcelona, Mart\'i i Franqu\`es 1, E08028 Barcelona, Spain}
    \affiliation{University of Barcelona Institute of Complex Systems (UBICS), Mart\'i i Franqu\`es 1, E08028 Barcelona, Spain}
    \affiliation{Computing and Understanding Collective Action (CUCA) Lab, Mart\'i i Franqu\`es 1, E08028 Barcelona, Spain}

\author{Ignacio Pagonabarraga}
    \affiliation{Departament de Física de la Matèria Condensada, Universitat de Barcelona, Mart\'i i Franqu\`es 1, E08028 Barcelona, Spain}
    \affiliation{University of Barcelona Institute of Complex Systems (UBICS), Mart\'i i Franqu\`es 1, E08028 Barcelona, Spain}

\date{\today}

\begin{abstract}
We experimentally, numerically and analytically explore the diffusive transport of active colloidal particles with sensory delay, navigating  motility landscapes in which the self--propulsion speed depends on space. We show how the transport properties can be obtained by replacing the space dependence of the self--propulsion speed by a dynamical stochastic switching process in the absence of delay, and extend the theory for systems with finite delayed responses. % to changes in the landscape. 
We obtain analytical results for the mean square displacement and the effective diffusion coefficient which accurately predict experimental measurements and numerical simulations across multiple scales. 
We show how, within the regime of validity of the delay-extended theory, density patterns and effective diffusion obey universal scaling forms. Our work provides minimal framework describing the transport properties of active swimmers with internal adaptation dynamics in motility landscapes.
\end{abstract}

\maketitle

% \section{Introduction} \label{sec:intro}
In active matter, non--trivial transport properties can emerge from the interplay between self-propulsion and environmental heterogeneity \cite{Bechinger16}. 
At the micrometer scale, suspensions of bacteria or Janus colloids provide well-controlled realizations of active agents moving through heterogeneous landscapes generated by nutrient, chemical, or light gradients \cite{van2004chemotaxis, cates2012diffusive, raynaud2014spatial, lozano2016phototaxis, arlt2018painting, frangipane18, fernandez2020feedback, massana2022rectification, Tpfer2025}.  While in thermal systems spatial heterogeneity typically leads to position-dependent diffusion coefficients and anomalous transport \cite{Fulinski13Communication, Cherstvy13, Cherstyv13Royal, Cherstyv14Physical, Cherstyv14Soft, Singh22}, in active systems it instead modulates the propulsion speed, giving rise to steady-state density and polarization patterns \cite{Auschra21, Auschra21curved, Holubec26}.  
Active swimmers in spatially varying propulsion fields have been extensively studied, mainly in terms of steady-state density and polarization patterns, showing a tendency to accumulate, or localize, in regions where they move faster \cite{Schnitzer93,Tailleur08,stenhammar16,sharma17, lozano19propagating, Auschra21, Auschra21curved,caprini22, caprini22external, vuijk22, Wysocki22, desgranges23, Nasiri23}. 
These descriptions usually assume instantaneous adaptation to the environment. However, in many systems propulsion responds with sensory delay \cite{Tpfer2025}. While recent work has examined how delay alters stationary patterns \cite{mijalkov16,leyman18,Holubec26}, its impact on transport properties, such as mean-square displacements and effective diffusion, remains largely unexplored.

In this Letter, we develop a minimal theoretical framework characterizing the dynamics of self-propelled Janus colloids moving in light-induced motility landscapes. The theory yields quantitative analytical predictions in good agreement with experiments on active colloidal particles navigating checkerboard motility patterns in the presence of sensory delay.
We show that the dynamics can be recast as an effective stochastic switching of the propulsion speed, described by a telegraph process \cite{Datta24, Santra24}. This perspective decouples the environmental modulation from the particles' self-propulsion, providing a simple route to capture transport properties analytically, and of introducing finite sensory delays in a controlled manner. 
We establish a direct link between static and dynamic properties, showing that stationary density patterns and  effective diffusion obey universal scaling laws quantitatively captured by the theory and validated by experiments and simulations. The presence of sensory delay %, this structure \textcolor{blue}{is preserved but renormalized}
leads to non-monotonic and optimizable particle localization in low--motility regions and diffusive transport, with quantitative agreement across a broad range of parameters. The resulting description thus provides a general framework for active transport in heterogeneous environments.

\begin{figure*}[t!]
    \centering
    \includegraphics[width=0.99\textwidth]{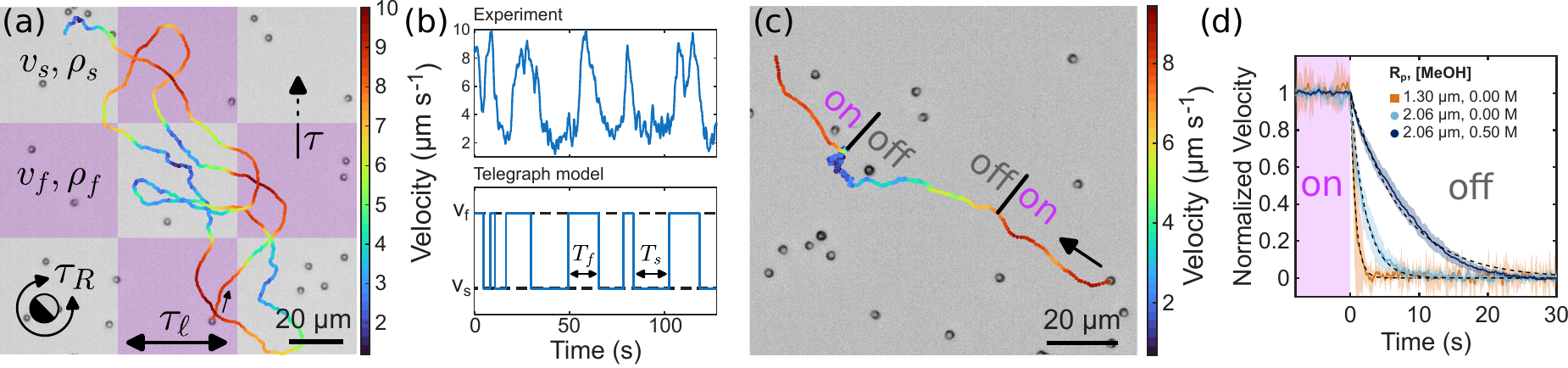}
    \vspace{-0.2cm}
\caption{
%\textbf{Checkerboard motility landscape and temporal response.}
\textbf{(a)} Experimental microscopy image with an overlaid single-particle trajectory at 8\,V, 4\,kHz and illumination intensity \SI{6.3}{\watt\per\centi\meter\squared}.
Color map indicates instantaneous speed; arrow denote propulsion direction.
Violet regions mark illuminated checkerboard domains.
\textbf{(b)} \emph{Top:} Propulsion speed \(v(t)\) in the checkerboard landscape.
\emph{Bottom:} Two-state telegraph process switching between \(v_f\) and \(v_s\), reproducing the measured speed statistics via mean occupation times \(\langle T_f \rangle\) and \(\langle T_s \rangle\).
\textbf{(c)} Single-particle trajectory under global temporal illumination switching (on–off–on, with \SI{6.3}{\watt\per\centi\meter\squared}), at 4\,V, 4\,kHz and 0.5\,M methanol; arrow indicates the direction of motion.
\textbf{(d)} Normalized velocity decay following a global illumination switch, which we used to extract the delay time in the experiments. The delay time can be experimentally varied by adding methanol. Solid lines show ensemble averages; shaded areas indicate the 95\% confidence interval of the mean (standard error from at least three independent segments).}
\label{fig:checkerboard_timescales}
\end{figure*}

\paragraph{Experimental setup.} 
Our experimental platform is built around UV-responsive silica--titania Janus colloids that self-propel under uniform AC electric fields via induced-charge electrophoresis (ICEP) \cite{SQUIRES2006,SQUIRES2004}. 
UV illumination increases the conductivity of the \ce{TiO2} cap \cite{Zehavi2022,Tpfer2025}, enabling spatial control of the propulsion speed and the realization of motility landscapes.
In the checkerboard pattern of Fig.~\ref{fig:checkerboard_timescales} (a) made of identical square domains of area $\ell\times\ell$, particles move with two well-defined speeds, a faster $v_f$ and a slower $v_s$, in illuminated and dark regions, respectively. Repeated crossings between domains %make the instantaneous
lead to a speed switch between these two values, yielding a two-level time series [see Fig.~\ref{fig:checkerboard_timescales}(b)], with stochastic residence times. This behaviour motivates an effective two-state telegraph description, fully parameterized by the measured mean residence times $\langle T_f\rangle$ and $\langle T_s\rangle$ in the fast and slow region, respectively.

Beyond imposing a spatially modulated self-propulsion speed, $v(\mathbf r)$, the experimental setup also enables control of sensory delay. Following a step change in illumination, the propulsion speed relaxes exponentially, see Fig. ~\ref{fig:checkerboard_timescales}(c,d)--, defining a \emph{delay time $\tau$}. By adding methanol, which suppresses charge-carrier recombination in the \ce{TiO2} cap \cite{Mohamed2011,Schneider2014,Tpfer2025}, $\tau$ can be tuned independently of the landscape geometry, from nearly instantaneous adaptation to pronounced sensory delay.
This makes the system a controlled testbed to probe both structure and transport in motility landscapes.
%Localization is quantified by the occupation fraction $\eta$,  measured from the steady-state particle distribution across fast and slow domains, while dynamics are captured by the mean-square displacement and the long-time effective diffusion coefficient $D_{\mathrm{eff}}$. Together, the independently tunable parameters $\{v_f,v_s\}$, $\tau$ and the landscape length scale $\ell$,  allow us to directly connect steady-state localization to diffusive transport. 
Experimental details and analysis procedures are provided in the Supplemental Material (SM) \cite{SM} and Ref.~\cite{Tpfer2025}.

\paragraph{Model.} 
Models for  self--propelled particles in which the activity is not homogeneously distributed in space have been presented in \cite{Auschra21, Auschra21curved, caprini22, caprini22external, vuijk22}. %Typically, the presence of activity landscapes is modeled through the active Brownian particle (ABP) model in two dimensions, governed by the following Langevin equations
Typically, the presence of activity landscapes is modeled within the active Brownian particle (ABP) model, where the self--propulsion velocity is a space-dependent function, i.e. $v = v(\textbf{r})$. In this case, the equations of motion of an ABP in two dimensions become
\begin{align}\label{eq:eom no delay}
    &\displaystyle\dot{\textbf{r}}_t = v(\textbf{r}_t)\hat{\nu}(\theta_t) + \sqrt{2D}\boldsymbol{\eta}(t) , \quad \displaystyle\dot{\theta}_t = \sqrt{2D_r} \chi(t) 
\end{align}
%where the self-propulsion velocity $v$,  is  spatially dependent.  Here,  
where $\hat{\nu}(\theta) = (\cos \theta, \sin \theta)^T$ is the ABP's director, $D$ and $D_r$ are the translational and rotational diffusion coefficients respectively, and  $\boldsymbol{\eta}(t)$ and $\chi(t)$ are independent, unitary, Gaussian white noises.

A moment expansion of the Fokker--Planck equation associated with Eqs.~(\ref{eq:eom no delay}), together with a closure ansatz at the level of the first two moments of the associated probability density function, i.e. the density $\rho(\textbf{r},t)$ and the polarization $\textbf{p}(\textbf{r},t)$ fields, may be used analytically access stationary density and polarization patterns induced by the landscape $v(\textbf{r})$ \cite{Auschra21,Auschra21curved}.
When the domain size $\ell$ is much larger than the characteristic length scales $\lambda_{f,s}=\sqrt{D/D_r}/\sqrt{1+\mathrm{Pe}_{f,s}}$ \cite{Auschra21}, where $\mathrm{Pe}_{f,s}=v_{f,s}^2/(2DD_r)$ are the Péclet numbers of fast and slow regions, %i.e. when $\lambda_{f,s}/\ell \ll 1$, 
the bulk densities can be approximated by the average densities in each region. For an ensemble of $N$ non--interacting ABPs traveling across the landscape, the average densities are provided by the number of particles in slow, $n_s$, and fast, $n_f$, regions. It follows that \cite{Auschra21}
\begin{align}\label{eq:hydrodynamic localization}
    \frac{\eta}{1-\eta} = \sqrt{\frac{1 + \mathrm{Pe}_f}{1 + \mathrm{Pe}_s}},
\end{align}
where $\eta = \frac{n_s}{N}$ and $1 - \eta = \frac{n_f}{N}$ represent the \textit{occupation fraction} of slow and fast regions respectively.

\paragraph{Telegraph process approach.}
While the moment expansion of the Fokker--Planck equation captures the stationary patterns induced by the landscape $v(\mathbf r)$, it does not directly address transport. Here the key difficulty is that the propulsion speed itself becomes a stochastic process, as it depends on the particle trajectory and therefore on both orientational diffusion and thermal noise. As illustrated in Fig.~\ref{fig:checkerboard_timescales}(b), a single trajectory then yields a two-level time series in which the speed switches between  $v_f$ and $v_s$ with random residence times.
Such random switching  has  been recently considered in  \cite{Datta24, Santra24} to understand intermittent behavior and its emergent transport properties in the ABP model, although in our case the switching is intrinsically coupled to space.

To obtain analytical access to the dynamics, we replace the explicit spatial dependence by an effective telegraph process, \(v(\mathbf r_t)\to v_t\), where \(v_t\) switches between \(v_f\) and \(v_s\) with rates \(\alpha=\langle T_s\rangle^{-1}\) and \(\beta=\langle T_f\rangle^{-1}\), see Fig. \ref{fig:checkerboard_timescales}(b). 
In this description, the landscape enters only through the switching statistics, and the stationary probabilities of the two states are directly identified with the \textit{occupation fractions} of slow and fast regions, i.e. the stationary probabilities are $\eta = \frac{\beta}{\alpha+\beta}$ and $1- \eta = \frac{\alpha}{\alpha+\beta}$ respectively. These values represent the \textit{fraction} of realizations for which one finds an ABP with a given self--propulsion velocity in steady conditions. %In a landscape, we directly identify these two as the \textit{occupation fraction} of fast and slow domains, $1-\eta$ and $\eta$. %The MSD, in this case, can be computed the following way,

As shown in the SM \cite{SM} the corresponding mean square displacement (MSD) can be computed analytically  assuming  that the rotational and telegraphic noise are independent. It reads, 
\begin{widetext}
\begin{align}\label{eq:MSD closed}
    \text{MSD}(t) = 4Dt + 2\overline{v}^2\left[\frac{t}{D_r} -\frac{1}{D_r^2}(1-e^{-D_r t})\right] +2(1-\eta)\eta (v_f-v_s)^2\left[\frac{t}{D_r(1+\delta)} - \frac{1}{D_r^2(1+\delta)^2}(1-e^{-D_r(1+\delta)t})\right],
\end{align}
\end{widetext}
where $\bar{v} = \eta v_s + (1-\eta)v_f$ and $\delta = \frac{\langle T_s\rangle^{-1} + \langle T_f\rangle^{-1}}{D_r}$.  
Contrary to ABPs, the MSD is characterized by  two  time--scales $D_r^{-1}$ and  $D_r^{-1}(1+\delta)^{-1}$. %While the term only including $D_r$ in Eq. (\ref{eq:MSD closed}) corresponds to the MSD of an ABP with velocity $\bar{v}$, the timescale $D_r^{-1}(1+\delta)^{-1}$ features the presence of the landscape and the dynamic switching between fast and slow states. 
While the term in Eq. (\ref{eq:MSD closed}) that only includes $D_r$ corresponds to the MSD of an ABP with velocity $\bar{v}$, the timescale $D_r^{-1}(1+\delta)^{-1}$ features the presence of the landscape and the dynamic switching between fast and slow states. The corresponding  effective diffusion coefficient  thus reads
\begin{align}\label{eq:eff diff coeff general}
    D_{\text{eff}} = D + \frac{1}{2D_r}\left[\bar{v}^2 + \frac{(1-\eta)\eta (v_f-v_s)^2}{1 + \delta}\right].
\end{align}
In the limit of slow switching, \(\delta\ll 1\), this reduces to the scaling form
\begin{align}\label{eq:scaling eff diff coeff}
    \frac{D_{\text{eff}}}{D}\frac{1}{1+\text{Pe}_s} = \frac{\eta}{1-\eta} + \mathcal{O}\left(\delta\right).
\end{align}
showing that the long-time diffusion is directly controlled by the steady-state localization.  It is important to note that quantities such as  the MSD in Eq. (\ref{eq:MSD closed}) and $D_{\text{eff}}$ in Eq. (\ref{eq:eff diff coeff general}) involve averages over the self-propulsion velocity distribution $p(v) = \eta \delta(v-v_s) + (1-\eta)\delta(v-v_f)$. As shown in the SM \cite{SM}, when $\delta \ll 1$  the effective diffusion coefficient is approximately $D_{\text{eff}}/D \approx 1 + \overline{\text{Pe}}$, where  $\text{Pe} = v^2/2DD_r$ and the average $\overline{\circ}$ is done over  $p(v)$ \cite{applicability_th_delay}. 
\begin{figure*}[t]
    \centering
    \includegraphics[width=0.99\textwidth]{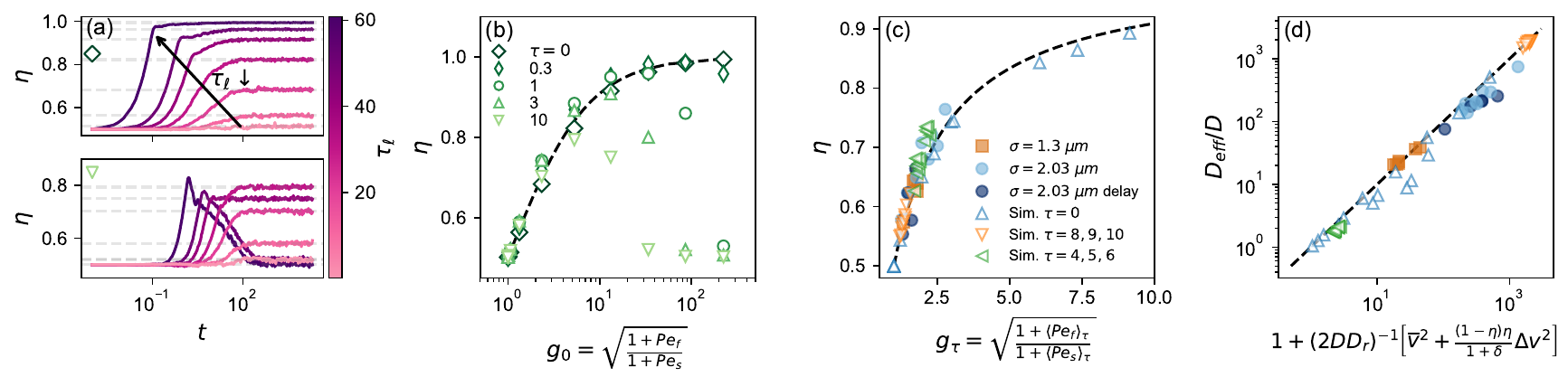}
    \vspace{-0.5cm}
    \caption{
    %\textbf{Effect of sensory delay on localization and effective diffusion coefficient.} 
    \textbf{(a)} Time evolution of the occupation fraction $\eta(t)$ %in a system of $N = 10^4$ non--interacting ABPs 
    in a checkerboard landscape of total size $L = 1$ and domain size $\ell = 1/4$ with $v_s = 0$ and $v_f = \ell/\tau_{\ell}$, with $\tau_{\ell} \in [0.1, 60.8]$ and for $\tau_r = 1$, $\tau_D = 10^3$ and $\tau = 0$ (top), $\tau = 10$ (bottom). \textbf{(b)} Stationary values of the occupation fraction $\eta$ in the same conditions as in (a) and different values of $\tau$. The dashed line represents the theoretical prediction in the $\tau=0$ limit. \textbf{(c)} Collapse of the stationary values of  $\eta$ in the presence of delay using the scaling in Eq. (\ref{eq:localization with delay}). Circles represent experiments performed with particles of $2.03\;\mu$m of diameter without and with methanol doping \cite{SM} and squares with  particles of $1.3\;\mu$m of diameter, without methanol doping. The triangles represent simulations performed in the absence  (blue) and in the presence of delay (green, orange). \textbf{(d)} $D_{\text{eff}}$ corresponding to the  experiments and simulations  in panel (c). The dashed line represents the complete shape of the prediction  Eq. (\ref{eq:eff diff coeff general}), where $\alpha$ and $\beta$ are estimated using self--propulsion speed time series. Note that here $\Delta v^2 = (v_f-v_s)^2$.} %\dl{Figure improvements: add labels a, b, c... inside the figure. The labels are too small, make the figs bigger, there is a lot of white space.}
    \label{fig:delay localization and eff diff coefficient}
\end{figure*}

\paragraph{Sensory delay.} 
So far, the model has only considered an instantaneous response of the particles as they get into a \textit{slow} or \textit{fast} region. 
However, particles generally possess intrinsic finite adaptation times to any change in the environment. %As shown in Fig \ref{fig:checkerboard_timescales}(c, d), in this particular setup particles exhibit an exponential relaxation to en. We incorporate such a time dependence of the self-propulsion velocity, that now obeys 
In our experimental system, the transition between fast and slow states is well described by an exponential function, controlled by the delay time $\tau$. This can be modeled transforming the self--propulsion velocity into a time--dependent object obeying
\begin{align}\label{eq:sensory delay sp velocity}
    \dot{v}(\textbf{r}_t, t)= - \tau^{-1}[v(\textbf{r}_t, t) - u(\textbf{r}_t)],
\end{align}
where  $u(\textbf{r})$ is the stationary value of the velocity set by the motility landscape or \textit{local velocity target}. The presence of sensory delay through Eq. (\ref{eq:sensory delay sp velocity}) results in non--Markovian translational dynamics, where now an exponential memory kernel shapes the self--propulsion velocity of the particle, see SM \cite{SM}. 

The sensory delay strongly modifies the patterns induced by the motility landscape. In the large-delay limit, the density distribution becomes homogeneous, whereas at intermediate delays the occupation fraction can display a non-monotonic response and even reach an optimum depending on the landscape parameters \cite{Tpfer2025}. As shown in the SM \cite{SM}, the dynamics of a sensory-delayed active colloid can be mapped onto the delay-free dynamics using Eq.~(\ref{eq:scaling eff diff coeff}) as an \textit{ansatz} and replacing the bare Péclet numbers  $\mathrm{Pe}_{f,s}$  by delay-renormalized effective values, $\overline{\mathrm{Pe}_{f,s}^{\tau}}$, obtained from the self-propulsion speed distribution. As a result,%, up to order $D_r^{-1}$. Thus,
\begin{align}\label{eq:localization with delay}
    \frac{\eta_{\tau}}{1-\eta_{\tau}} = \sqrt{\frac{1+\overline{\text{Pe}_f^{\tau}}}{1+\overline{\text{Pe}_s^{\tau}}}},
\end{align}
where the $\tau$ subscript indicates that $\eta_{\tau}$ and $1-\eta_{\tau}$ are the occupation fractions in the presence of delay. Furthermore, similarly to Eq. (\ref{eq:scaling eff diff coeff}),
\begin{align}\label{eq:scaling D_eff delay}
    \frac{D_{\text{eff}}}{D}\frac{1}{1+\overline{\text{Pe}_s^{\tau}}} = \frac{\eta_{\tau}}{1-\eta_{\tau}} + \mathcal{O}(D_r^{-1}).
\end{align}
Although the probability density function in this case, $p_{\tau}(v)$, is not explicitly known, $\overline{\mathrm{Pe}}_{f,s}$ can be numerically and experimentally measured, see SM \cite{SM}. Note that while Eq.~(\ref{eq:scaling D_eff delay}) is a consequence of the \textit{ansatz} itself, the scaling in Eq.~(\ref{eq:localization with delay}) -- which is the same scaling law as Eq.~(\ref{eq:hydrodynamic localization}) -- is a necessary conditions for the \textit{ansatz} to hold. This signals the presence of the same universal scaling laws, both in the case with and without delay. 

The details of the simulations and the implementation of the time--delayed responses introduced in Eq.~(\ref{eq:sensory delay sp velocity}) can be found in Refs.~\cite{fernandez2020feedback, Bailey2024, Tpfer2025}.

%\paragraph{Results .---} 

As shown in Fig. \ref{fig:delay localization and eff diff coefficient}(b), the occupation fraction becomes non-monotonic in the presence of sensory delay,  and it can be optimized for specific landscape parameters, while its maximum decreases with increasing $\tau$, approaching the homogeneous limit $\eta=1/2$ at large delay. The dashed line displays the occupation fraction in the absence of delay, which accurately describes the localization in the presence of delay until the optimal value is reached. The corresponding scaling collapse, shown in Fig.~\ref{fig:delay localization and eff diff coefficient}(c), confirms that the delay-renormalized effective P\'eclet numbers accurately capture the data. An explicit explanation of how effective Péclet numbers are measured can be found in the SM \cite{SM}.

\begin{figure}[h!]
    \centering
    \includegraphics[width=0.99\linewidth]{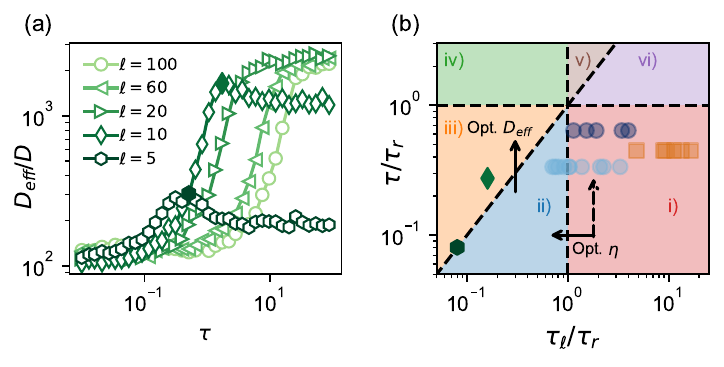}
    \vspace{-0.6cm}
    \caption{
    %\textbf{Non--monotonic behaviour of the effective diffusion coefficient and scale separation.} 
    (a) Effective diffusivity as a function $\tau$ for  $v_f = 10, v_s = 1$,  $\tau_r = 6.26$, $D = 1.25\times10^{-12}$, system size $L = 10^3$ and domain sizes $\ell$ ranging from $10$ to $100$. $D_{\text{eff}}$ either grows monotonically with delay or shows an optimal value at given $\tau^*$ for $\ell \leq 10$. (b) Time--scale separation of the experiments and the optimal value for $\ell = 10$ in panel a). The latin numerals indicate all the dynamic (ordering) regimes of the time scales $\tau_r = 1/D_r$, $\tau_{\ell} = \ell/v_f$ and $\tau$. Note that only the fast regions are represented, while corresponding slow regions of each experiment lay on the right of each scatter. The green diamond and hexagon shaped--markers correspond to the maxima in panel a).} 
    \label{fig:non_monot_D_eff_and_separation_of_scales}
\end{figure}

Fig.~\ref{fig:delay localization and eff diff coefficient}(d) displays the performance of the theoretical framework to reproduce $D_{\text{eff}}$ measured in experiments and simulations.
%The same framework also quantitatively describes the effective diffusion coefficient measured in experiments and simulations, see Fig.~\ref{fig:delay localization and eff diff coefficient}(d). 
Measurements of the effective diffusion coefficient (markers) are plotted along the prediction Eq.~(\ref{eq:eff diff coeff general}) (dashed line), where $\delta$ is estimated through measurements of the average residence time in slow and fast regions using individual trajectories.  The results show the ability of Eq.~(\ref{eq:eff diff coeff general}) to reproduce the transport behavior across scales, both with and without delay.  Beyond this agreement, the framework reveals that both $\eta$ and $D_{\mathrm{eff}}$ can vary non-monotonically with $\tau$ -- for fixed landscape parameters, both localization $g = \frac{\eta}{1-\eta}$~\cite{Tpfer2025} and the effective diffusion coefficient can be optimized at a finite delay time, see Fig.~\ref{fig:non_monot_D_eff_and_separation_of_scales}(a).

The results in Figs.~\ref{fig:delay localization and eff diff coefficient} and \ref{fig:non_monot_D_eff_and_separation_of_scales} show that the model accurately predicts the effective diffusion coefficient both with and without delay. In the delay-free case, the prediction in Eq.~(\ref{eq:eff diff coeff general}) applies to all $\delta$ when the domain size satisfies $\ell \gg \lambda_{f,s}=\sqrt{D/D_r}/\sqrt{1+\mathrm{Pe}_{f,s}}$, so that the bulk relation in Eq.~(\ref{eq:hydrodynamic localization}) applies. In the presence of delay, the theory is valid in the slow-switching regime $\delta=(\langle T_s\rangle^{-1}+\langle T_f\rangle^{-1})/D_r \ll 1$, which requires the mean residence times to be much longer than the persistence time $\tau_r=1/D_r$.

As shown in Fig.~\ref{fig:non_monot_D_eff_and_separation_of_scales}(b), all the experimental realizations fall within these regimes i) and ii), for which, even in the fast domains, where residence times are shortest, we find $\tau_\ell/\tau_r \gtrsim 1$, implying that $\langle T_s\rangle^{-1}/D_r<1$ and $\langle T_f\rangle^{-1}/D_r<1$. This confirms that the required separation of timescales is satisfied experimentally. The delay-extended theory therefore provides reliable predictions in the corresponding parameter regimes, while deviations observed outside this limit point to the need for a framework able to treat delay exactly at arbitrary switching rates. The dashed and solid arrows in 
Fig.~\ref{fig:non_monot_D_eff_and_separation_of_scales}(b) further depict the 
phenomenological behavior when traversing these regimes, including the 
optimization of $\eta$, e.g. crossing from i) to ii), and $D_{\mathrm{eff}}$, e.g crossing from ii) to iii).  A more detailed explanation of the different regimes i) - vi) with a descriptive table going through phenomenological behavior and the applicability of the delay-extended theory can be found in the SM \cite{SM}.

\paragraph{Conclusion.} 
We have shown that the long-time dynamics and transport properties of an active colloid in a heterogeneous landscape, where the self-propulsion speed depends on position, can be accurately described by replacing the explicit spatial modulation with a continuous-time stochastic process that preserves the switching statistics imposed by the landscape. This effective description remains valid both in the absence and presence of finite sensory delay. Sensory delay modifies the distribution of self-propulsion speeds and thereby renormalizes the resulting transport properties. Remarkably, within the regime of validity of the theory, the underlying structure remains unchanged, so that localization and effective diffusion still obey the same universal scaling forms. %The resulting framework provides a minimal description of transport in active systems with internal adaptation dynamics, likely to be relevant to both biological and synthetic swimmers.
As reported in \cite{Tpfer2025}, the sensory delay present in the experimental system is asymmetric; particles relax with a finite
delay time when transition from fast to slow regions, while the reverse transitions happens much faster. Since the model presented in this manuscript only considers a single response timescale, asymmetric responses can be considered in future work.
The resulting framework establishes a minimal description of transport in active systems with internal adaptation dynamics, connecting swimmer-level stochastic processes to macroscopic transport in a way that may prove useful across a broad class of non-equilibrium systems.

\paragraph{Acknowledgments .--} L.I. and U.T. acknowledge funding from the European Research Council (ERC) under the European Union’s Horizon 2020 Research and Innovation Program grant agreement No. 101001514. The authors also acknowledge the use of cleanroom facilities at the Binnig and Rohrer Nanotechnology Center (BRNC) at IBM Research – Zurich and at the FIRST Center for Micro- and Nanoscience at ETH Zurich.  A.G. acknowledges AGAUR and Generalitat de Catalunya for financial support under the call FI SDUR 2023 Ref.~CCI 2021ES05FPR011. D.L. acknowledges MCIU/AEI for financial support under grant agreement PID2022-140407NB-C22. I.P acknowledges support from Ministerio de Ciencia, Innovaci\'on y Universidades MCIU/AEI/FEDER for financial support under grant agreement PID2024-156516NB-100 AEI/FEDER-EU, and Generalitat de Catalunya for financial support under Program Icrea Acad\`emia and project 2021SGR-673.

\paragraph{Data availability .---} The data is available from the authors upon reasonable request.

\bibliography{references.bib}% Produces the bibliography via BibTeX.

\begin{widetext}

\onecolumngrid
\vspace{2em}
\begin{center}
  \textbf{\large Supplemental Material of ``Universal transport of active colloids with sensory delay in motility landscapes''}
  \vspace{1em}

  Adri\`a Garc\'es,\textsuperscript{~1,2,3,$*$} Ueli Töpfer,\textsuperscript{~4,$\dagger$} Lucio Isa,\textsuperscript{~4}  Demian Levis,\textsuperscript{~1,2,3} and Ignacio Pagonabarraga,\textsuperscript{~1,2} \\
 \vspace{0.2em}

  \textsuperscript{1}\textit{\small Departament de Física de la Matèria Condensada, Universitat de Barcelona, Mart\'i i Franqu\`es 1, E08028 Barcelona, Spain} \\
  \textsuperscript{2}\textit{\small University of Barcelona Institute of Complex Systems (UBICS), Mart\'i i Franqu\`es 1, E08028 Barcelona, Spain} \\
  \textsuperscript{3}\textit{\small Computing and Understanding Collective Action (CUCA) Lab, Mart\'i i Franqu\`es 1, E08028 Barcelona, Spain} \\
  \textsuperscript{4}\textit{\small Department of Materials, ETH Zurich, 8049 Zurich, Switzerland}
\end{center}
\vspace{1em}

\begingroup
\renewcommand\thefootnote{}\footnotetext{%
\href{mailto:adria.garces@ub.edu}{\textsuperscript{$*$} adria.garces@ub.edu}\\%\quad
\href{mailto:ueli.toepfer@mat.ethz.ch}{\textsuperscript{$\dagger$} ueli.toepfer@mat.ethz.ch}%
}
\endgroup

\setcounter{section}{0}
\setcounter{equation}{0}
\setcounter{figure}{0}
\setcounter{table}{0}
\renewcommand{\thesection}{S\arabic{section}}
\renewcommand{\theequation}{S\arabic{equation}}
\renewcommand{\thefigure}{S\arabic{figure}}
\renewcommand{\thetable}{S\arabic{table}}

\section*{I. Experimental Details and Materials}

\subsection*{A. Particle Fabrication and Experimental Setup}

Silica--titania Janus particles were fabricated from monodisperse silica spheres (diameter \SIlist{1.3;2.06}{\micro\meter}, microParticles GmbH). A dilute suspension (\SI{0.5}{\wv}) was deposited onto plasma-cleaned glass substrates and dried to form a colloidal monolayer. The exposed hemispheres were coated with a \SI{50}{\nano\meter} layer of \ce{Ti3O5} using an e-beam evaporator (Plassys MEB700SL). The samples were annealed at \SI{500}{\celsius} for \SI{120}{\minute} under \SI{1}{\bar} \ce{O2} in a rapid thermal processor (AS-One~150, Annealsys), producing predominantly anatase \ce{TiO2}. The particles were released by sonication and redispersed in Milli-Q water.

Experiments were performed in sealed chambers consisting of two ITO-coated glass slides (sheet resistance \(<\SI{7}{\ohm\per\square}\), Präzisions Glas \& Optik GmbH) separated by a \SI{120}{\micro\meter} adhesive spacer (SecureSeal, Grace Bio-Labs, USA). The spacer defined a circular chamber of \SI{9}{\milli\meter} diameter with inlet and outlet channels of approximately \SI{1}{\milli\meter} width. \SI{9.5}{\micro\liter} of particle suspension was injected into the chamber.

Prior to loading, the particle suspension was mixed with a Pluronic F-127 (Sigma-Aldrich) stock solution to obtain a final surfactant concentration of either \SI{0.1}{\wv} or \SI{0.3}{\wv}, reducing particle adhesion to the substrate. The channels were sealed with Krytox GPL205 grease to minimize evaporation. Electrical contact was made using adhesive copper tape, and uniform AC electric fields were applied using a function generator (Agilent 33522A) with amplitudes between \SIrange{1}{10}{\volt} and frequencies of \SIrange{4}{10}{\kilo\hertz}.

UV illumination at \SI{365}{\nano\meter} was provided by a Prizmatix UHP-F-365 LED and spatially patterned using a digital micromirror device (Polygon~1000, Mightex). Reflected UV light was suppressed using a Nikon DAPI filter cube. Imaging was performed on a Nikon Eclipse Ti2-E inverted microscope using ELWD 40$\times$ and 60$\times$ objectives. Image sequences (2160\,\(\times\)\,2560\,px) were recorded at \SI{10}{\fps} using an Andor Zyla sCMOS camera. In selected experiments, methanol (\(\geq\SI{99.9}{\percent}\), Sigma-Aldrich) was added at a concentration of \SI{0.5}{\molar} to increase the sensory delay time.

\subsection*{B. Particle Tracking and Transport Parameters}

Particle positions were extracted from image sequences using custom MATLAB routines, and trajectories were reconstructed by frame-to-frame linking. Mean-squared displacements (MSDs) were computed from the tracked trajectories and ensemble-averaged over particles and time origins. Translational diffusion coefficients $D$ were determined independently under zero applied electric field, where particle motion was purely Brownian. In this regime, MSDs were fitted to
\begin{equation}
\label{eq:msd_brownian}
    \mathrm{MSD}(\Delta t) = 4 D \, \Delta t.
\end{equation}

The resulting diffusion coefficients for different particle sizes and methanol concentrations are summarized in Table~\ref{tab:diffusion}. Representative MSD curves and corresponding linear fits are shown in Fig.~\ref{fig:msd}(a), demonstrating purely Brownian diffusion in the absence of applied electric fields.

\begin{table}[h!]
\caption{Translational diffusion coefficient $D$ (in \si{\micro\meter\squared\per\second}) for particles of radius $R$ in the absence and presence of methanol. Values were extracted from fits to the mean-squared displacement using Eq.~\ref{eq:msd_brownian}. Uncertainties represent the standard error of the mean, estimated from fits to upper and lower bounds of the MSD standard error. The combination marked ``--'' was not investigated.}
\label{tab:diffusion}
\centering
\begin{tabular}{c|cc}
 & \multicolumn{2}{c}{Added methanol (M)} \\
$R\;(\mu\mathrm{m})$ & 0.0 & 0.5 \\
\hline
0.65 & $0.201 \pm 0.006$ & -- \\
1.03 & $0.11 \pm 0.02$ & $0.108 \pm 0.008$ \\
\end{tabular}
\end{table}

Rotational diffusion coefficients were fixed to their theoretical Stokes--Einstein values for spherical particles,
\begin{equation}
    D_r = \frac{k_{\mathrm{B}} T}{8\pi \eta R^3},
\end{equation}
yielding $D_r = \SI{0.636}{\radians^2\per\second}$ for particles of diameter \SI{1.3}{\micro\meter} and $D_r = \SI{0.160}{\radians^2\per\second}$ for particles of diameter \SI{2.06}{\micro\meter}.

Sensory delay times $\tau$ were extracted from exponential fits to transient propulsion velocities following global illumination ON/OFF cycles, as shown in Fig. 1 in the main text. The extracted delay times for different particle sizes and methanol concentrations are listed in Table~\ref{tab:delay}.

\begin{figure}[h!]
    \centering
    \includegraphics[width=0.4\columnwidth]{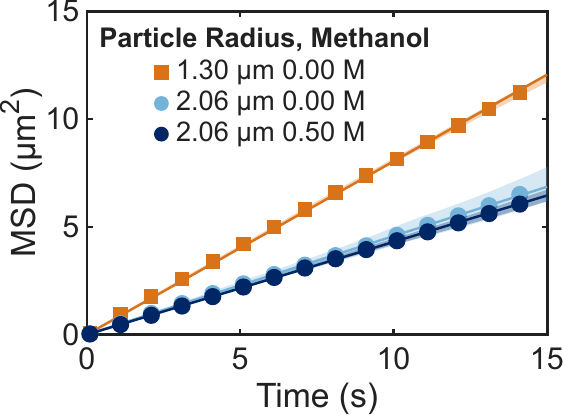}
    \caption{\textbf{Diffusion of passive particles.}
    Mean-squared displacement (MSD) of passive particles measured under zero applied electric field for particles of different diameters. Solid lines indicate linear fits to Eq.~\ref{eq:msd_brownian} used to extract the translational diffusion coefficient $D$. 
    Shaded regions denote the 95\% confidence interval of the mean MSD.}
    \label{fig:msd}
\end{figure}

\begin{table}[h!]
\caption{Sensory delay time $\tau$ (in seconds) extracted from exponential fits to velocity transients during global illumination ON/OFF cycles. Uncertainties denote the 95\% confidence interval of the mean over at least three cycles. The combination marked ``--'' was not investigated.}
\label{tab:delay}
\centering
\begin{tabular}{c|cc}
 & \multicolumn{2}{c}{Added methanol (M)} \\
$R\;(\mu\mathrm{m})$ & 0.0 & 0.5 \\
\hline
0.65 & $0.7 \pm 0.3$ & -- \\
1.03 & $2.1 \pm 0.5$ & $7.4 \pm 0.8$ \\
\end{tabular}
\end{table}

In checkerboard experiments, particle trajectories were segmented according to the local illumination state to extract domain-specific transport parameters. Localization was quantified by the localization parameter $\eta$, defined as the fraction of particles residing in the low-motility (slow) domains in the steady state,
\begin{equation}
\eta = \frac{n_s}{n_f + n_s},
\end{equation}
where $n_s$ and $n_f$ denote the number of particles in the slow and fast regions, respectively. An illustrative example of the temporal evolution of $\eta$ from particle counts is shown in Fig.~\ref{fig:transport}(a). Steady-state values were obtained by averaging $\eta$ over the final \SI{100}{\second} of each experiment.

In accordance with a two-state telegraph model, particle motion in the checkerboard landscape was described as stochastic switching between fast and slow motility states. Particle trajectories were mapped onto a binary time series based on the particle position relative to the checkerboard pattern.

From the resulting state sequences, individual residence times in slow and fast regions, denoted $T_s^{(i)}$ and $T_f^{(i)}$, were extracted for all trajectories within the steady-state regime. The mean occupation times were then obtained as 

\begin{equation}
\langle T_s \rangle = \frac{1}{N_{sf}} \sum_i T_s^{(i)}, \qquad
\langle T_f \rangle = \frac{1}{N_{fs}} \sum_i T_f^{(i)} ,
\end{equation}

where $N_{sf}$ and $N_{fs}$ denote the total number of slow-to-fast and fast-to-slow transitions, respectively, across all trajectories. These mean occupation times determine the steady-state localization parameter,

\begin{equation}
\eta = \frac{\langle T_s \rangle}{\langle T_s \rangle + \langle T_f \rangle}.
\end{equation}
A comparison between the localization parameter obtained from direct particle counting and the occupation-time-based estimate for all experiments is shown in Fig.~\ref{fig:eta_localization}(b).

\begin{figure}[h!]
    \centering
    \includegraphics[width=0.8\columnwidth]{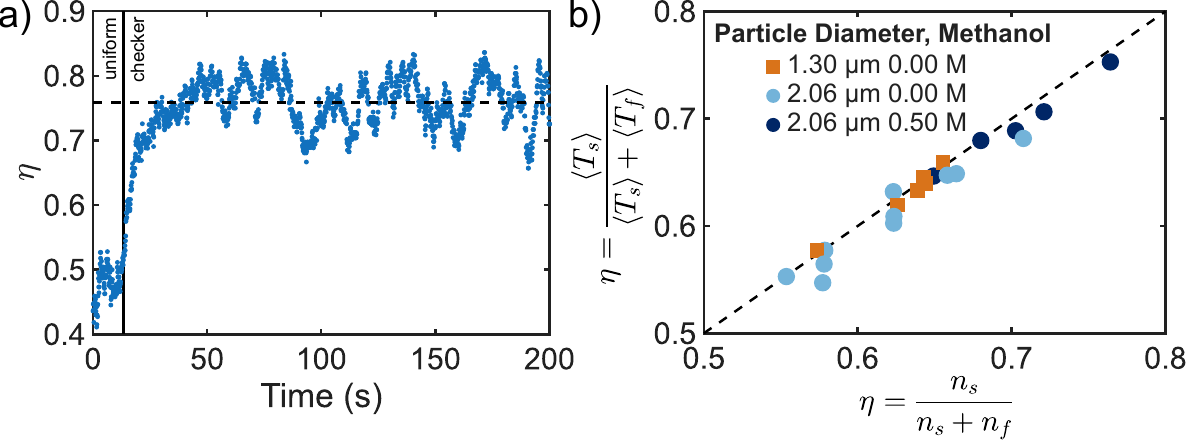}
    \vspace{-0.4cm}
    \caption{\textbf{Localization parameter in checkerboard motility landscapes.}
    (a) Temporal evolution of the localization parameter $\eta$ determined from particle counts in slow and fast regions; the vertical line indicates the transition from uniform illumination to checkerboard illumination, and the horizontal line denotes the mean over the final \SI{100}{\second} used to determine the steady-state value.
    Data are shown for a representative checkerboard experiment with a pattern period of \SI{40}{\micro\meter}, an applied electric field of \SI{8}{\volt} at \SI{4}{\kilo\hertz}, and an on-field UV intensity of \SI{6.3}{\watt\per\centi\meter\squared}.
\textbf{(b)} Comparison of steady-state localization parameters $\eta$ obtained from mean occupation times, $\eta=\langle T_s\rangle/(\langle T_s\rangle+\langle T_f\rangle)$, and from direct particle counting, $\eta=n_s/(n_s+n_f)$, for all checkerboard experiments.}
    \label{fig:eta_localization}
\end{figure}

Propulsion velocities in the fast and slow regions, $v_f$ and $v_s$, were extracted from ensemble-averaged mean-squared displacements computed from trajectory segments fully contained within the respective domains. The MSDs were fitted using the standard active Brownian particle (ABP) form \cite{Howse2007,Bechinger2016},
\begin{equation}
\label{eq:msd_active}
    \mathrm{MSD}(\Delta t) = \left(4D + 2\,v^2D_r^{-1}\right)\Delta t 
    + \frac{2\,v^2}{D_r^{2}}\left(e^{-\Delta t D_r}-1\right),
\end{equation}
where $v$ denotes the local propulsion speed and $D$ and $D_r$ were fixed to the independently determined values. Representative MSDs and corresponding fits used to extract $v_f$ and $v_s$ are shown in Fig.~\ref{fig:transport}(b).

The corresponding P\'eclet numbers were computed as
\begin{equation}
\mathrm{Pe}_{f,s} = \frac{v_{f,s}^2}{2 D D_r}.
\end{equation}

Long-time transport in checkerboard landscapes was characterized by an effective diffusion coefficient $D_\mathrm{eff}$, extracted from MSDs computed over full particle trajectories spanning multiple motility domains. At long lag times, $\Delta t \gtrsim D_r^{-1}$, the MSD was fitted to
\begin{equation}
\mathrm{MSD}(\Delta t) = 4 D_\mathrm{eff} \, \Delta t + C,
\end{equation}
where $C$ is a constant offset accounting for short-time dynamics. Representative MSDs and fits used to extract $D_\mathrm{eff}$ are shown in Fig.~\ref{fig:transport}(c).

\begin{figure}[h!]
    \centering
    \includegraphics[width=0.75\columnwidth]{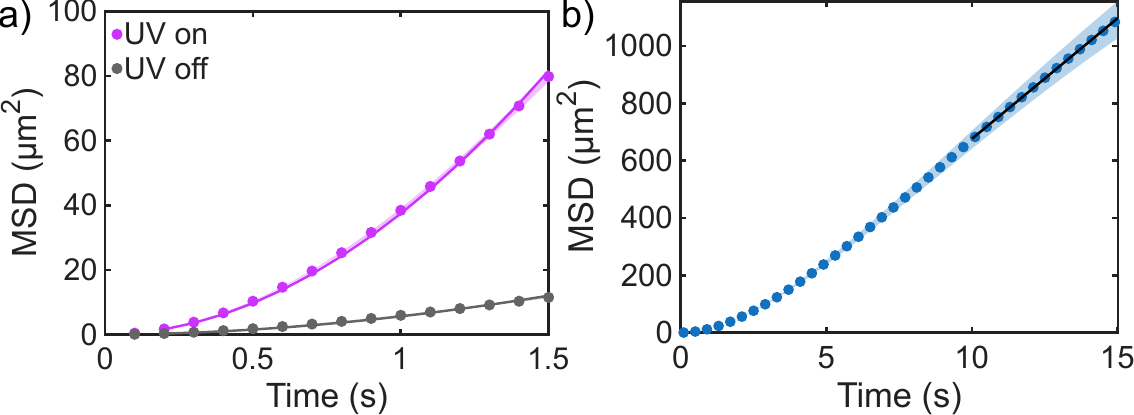}
    \vspace{-0.3cm}
    \caption{\textbf{Transport parameters in checkerboard motility landscapes.}
(a) Mean-squared displacements computed from trajectory segments confined to fast and slow regions (circles); solid lines indicate fits to Eq.~\ref{eq:msd_active} used to extract the local propulsion velocities $v_f$ and $v_s$, and shaded areas denote the standard error.
(b) Mean-squared displacements computed over full trajectories spanning multiple motility domains (circles); solid lines indicate linear fits used to extract the long-time effective diffusion coefficient $D_\mathrm{eff}$, and shaded areas denote the standard error.
Data shown correspond to the same representative checkerboard experiment as in Fig.~\ref{fig:eta_localization}(a).}
    \label{fig:transport}
\end{figure}

Figure~\ref{fig:residence} compares the mean residence times in the slow and fast regions with the rotational diffusion timescale $D_r^{-1}$ for all checkerboard experiments. 
For the majority of experiments, both $\langle T_s\rangle$ and $\langle T_f\rangle$ are comparable to or larger than $D_r^{-1}$, indicating predominantly diffusive reorientation within the motility domains. 
In particular, experiments performed with particles of smaller diameter \SI{1.30}{\micro\meter} predominantly fall in the diffusive regime in both motility states, reflecting their faster rotational dynamics.

\begin{figure}[h!]
    \centering
    \includegraphics[width=0.35\columnwidth]{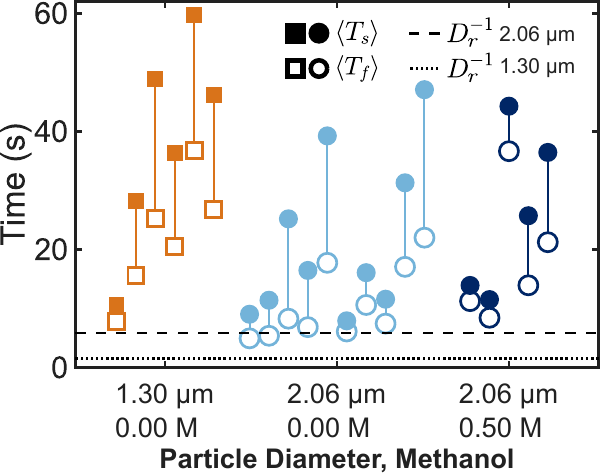}
    \vspace{-0.2cm}
 \caption{\textbf{Mean residence times and rotational timescale.}
Mean residence times $\langle T_s\rangle$ (filled symbols) and $\langle T_f\rangle$ (open symbols) measured in the slow and fast regions, respectively, for all checkerboard experiments, plotted relative to the rotational diffusion timescale $D_r^{-1}$. The comparison delineates the dynamical regime within motility domains: $\langle T\rangle \gg D_r^{-1}$ corresponds to diffusive reorientation, whereas $\langle T\rangle \lesssim D_r^{-1}$ indicates ballistic crossings.}
    \label{fig:residence}
\end{figure}

\newpage

\section*{II. MSD Calculation using the Telegraph Process Approach}
Consider an ABP model where now the self--propulsion velocity follows a telegraph process -- it stochastically transitions between two states with constant transition rates $\alpha$ and $\beta$. The model can be formalized by using the ABP model's equations of motion
\begin{align}
    \begin{cases}
        \displaystyle\dot{\textbf{r}}_t = v_t\hat{\nu}(\theta_t) + \sqrt{2D}\boldsymbol{\eta}(t) \\
        \displaystyle\dot{\theta}_t = \sqrt{2D_r} \chi(t),
    \end{cases}
\end{align}
where $v_t$ is the self--propulsion velocity running over a telegraph process, that is, transition between two states $v_t \in \{v_s,v_f\}$ (with $v_s < v_f$) with constant rates $\alpha$ and $\beta$. This means the velocity is a Markov process for which $p(v_{t+\Delta t} = v_f | v_{t} = v_s) = \alpha \Delta t$ while $p(v_{t+\Delta t} = v_s | v_{t} = v_f) = \beta \Delta t$, meaning the probabilities of finding the velocity at state $v_s$ or $v_f$ are governed, as mentioned in the main text, by the Master equation (ME)
\begin{align}
    \dot{p}_f(t) = \alpha p_s(t) - \beta p_f(t) \label{eq:me f}\\
    \dot{p}_s(t) = \beta p_f(t)-\alpha p_s(t) \label{eq:me s},
\end{align}
where we shorten the notation $p(v_t = v_{s,f})$ to $p_{s,f}(t)$. The general solution is obtained using conservation of probability, $p_s(t) + p_f(t) = 1$, giving
\begin{align}
    p_f(t|v_0,t_0) &= \frac{\alpha}{\alpha+\beta}+\left(p(v_0 = v_f) - \frac{\alpha}{\alpha+\beta}\right)e^{-(\alpha+\beta)(t-t_0)}\\
    p_s(t|v_0,t_0) &= \frac{\beta}{\alpha+\beta}+\left(p(v_0=v_s) - \frac{\beta}{\alpha+\beta}\right)e^{-(\alpha+\beta)(t-t_0)}\,.
\end{align}
These expressions depend on the initial distribution of the self--propulsion velocity. In the stationary state, the probabilities of finding the self--propulsion velocity $v_t$ to be $v_s$ and $v_f$ are $\beta/(\alpha+\beta) := \eta$ and $\alpha/(\alpha+\beta) := 1- \eta$ respectively. In order to simplify the problem, we will consider that the initial distribution of $v_t$ is the stationary one, so that the probabilities of the occupation of both state remain constant at any time $t$, $p_s(t) = \eta$, $p_f(t) = 1- \eta$. Note how directly from the ME one can integrate conditional probabilities, i.e. $p(v_{t} = v_{s,f} | v_{t'} = v_{s,f})$ for $t>t'$. In order to ease notation we will use the following to refer to the conditional probability, $p(v_{t} = v_{s,f} | v_{t'} = v_{s,f}) := p(v_{s,f},t | v_{s,f},t')$ for $t>t'$. The two-time solutions of the ME read
\begin{align}
    p_f(t|v_{t'},t') &= (1-\eta)+\left(p(v_{t'} = v_f) - (1-\eta)\right)e^{-(\alpha+\beta)(t-t')}\\
    p_s(t|v_{t'},t') &= \eta+\left(p(v_{t'}=v_s) - \eta\right)e^{-(\alpha+\beta)(t-t')},
\end{align}
where we have used $\eta = \beta/(\alpha+\beta)$, $1- \eta = \alpha/(\alpha+\beta)$. The conditional probabilities are thus simply given by
\begin{align}
    p(v_{f},t|v_{f},t') &:= p_f(t|v_{t'},t')|_{p(v_{t'}=v_f) = 1} =(1-\eta) + \eta e^{-(\alpha+\beta )(t-t')} \label{eq:cond-prob-f-f}\\
    p(v_{f},t|v_{s},t') &:= p_f(t|v_{t'},t')|_{p(v_{t'}=v_f) = 0} = (1-\eta)(1-e^{-(\alpha+\beta)(t-t')}) \\
    p(v_s, t | v_f, t') &:=p_s(t|v_{t'},t')|_{p(v_{t'}=v_s) = 0} = \eta(1-e^{-(\alpha+\beta)(t-t')}) \\
    p(v_s,t|v_s,t') &:= p_s(t|v_{t'},t')|_{p(v_{t'} = v_s) = 1} = \eta + (1-\eta)e^{-(\alpha+\beta)(t-t')}\,. \label{eq:cond-prob-s-s}
\end{align}
This can be used in order to compute the average displacement and the mean square displacement (MSD) of the model. Indeed, the average displacement of a particle up to time $t$ can be written as, 
\begin{align}\label{eq:avg displacement}
    \langle \Delta \textbf{r}_t\rangle := \langle \textbf{r}_t -\textbf{r}_0 \rangle = \int_{0}^{t} \langle v_{t'} \hat{\nu}(\theta_{t'}) \rangle \;dt' = \int_{0}^{t}\langle v_{t'}\rangle \langle \hat{\nu}(\theta_{t'}) \rangle\;dt'
\end{align}
since the telegraph process running on top of the self--propulsion velocity is independent of the noise realization of the angular and translational degrees of freedom. Now, because $\langle \hat{\nu}(\theta_{t})\rangle = \hat{\nu}(\theta_0)e^{-D_r t}$ and since
\begin{align}\label{eq:avg velocity}
    \langle v_{t'} \rangle := \sum_{ v_{t'} \in \{ v_s, v_f\}} v_{t'} p(v_{t'},t') = v_sp(v_s,t') + v_f p(v_f,t') = v_s \eta +v_{f} (1-\eta),
\end{align}
%since $p_{s}(t) = \eta$ and $p_f(t) = 1-\eta$ for all $t>0$. 
it follows that %Plugging this into Eq. (\ref{eq:avg displacement}) one finds,
\begin{align}
    \langle \Delta \textbf{r}_t \rangle = \bar{v} \hat{\nu}(\theta_0)\frac{1}{D_r}(1-e^{-D_r t}),
\end{align}
where in Eq. (\ref{eq:avg velocity}) we have used that $p_{s}(t) = \eta$ and $p_f(t) = 1-\eta$ for all $t>0$, and where we define $\bar{v} := \eta v_s + (1-\eta) v_f$. Note how this actually corresponds to a \textit{quench average} of the velocity over the velocity distribution $p(v) = \eta \delta(v-v_s) + (1-\eta)\delta(v-v_f)$, $\bar{v} = \int v p(v)\; dv$ . The same thing can be done  for the MSD. Directly from the model it is found that
\begin{align}\label{eq:supp msd}
    \langle \Delta \textbf{r}_t^2 \rangle := \langle (\textbf{r}_t-\textbf{r}_0)^2 \rangle = 4Dt + \int_{0}^{t}\int_{0}^{t}dt'dt''\;\langle v_{t'}v_{t''}\rangle \langle \hat{\nu}(\theta_{t'})\cdot \hat{\nu}(\theta_{t''})\rangle,
\end{align}
where we have again used that the telegraph process running on the self--propulsion velocity is independent of the rotational and translation noise realizations. Note how now the velocity correlations have to be computed for both cases $t'>t''$ and $t'<t''$, while $\langle \hat{\nu}(\theta_{t'})\cdot\hat{\nu}(\theta_{t''}) \rangle = e^{-D_r |t'-t''|}$. For $t'>t''$ we have,
\begin{align}
    \langle v_{t'}v_{t''} \rangle := \sum_{v_{t'},v_{t''} \in \{v_s,v_f\}} v_{t'}v_{t''} p(v_{t'},t';v_{t''},t'') = \sum_{v_{t'},v_{t''} \in \{v_s,v_f\}} v_{t'}v_{t''} p(v_{t'},t'|v_{t''},t'') p(v_{t''},t''),
\end{align}
where $p(v_{t'},t';v_{t''},t'')$ is the joint probability of finding some value of the self--propulsion velocity at times $t'$ and $t''$ and where in the last step we simply used Bayes' theorem. We use now the conditional probabilities Eq. (\ref{eq:cond-prob-f-f})--(\ref{eq:cond-prob-s-s}) and find
\begin{align}
    \langle v_{t'}v_{t''}\rangle &= \sum_{v_{t'}\in\{v_s,v_f\}}v_{t'} \sum_{v_{t''}\in\{v_s,v_f\}}v_{t''}p(v_{t'},t'|v_{t''},t'')p(v_{t''},t'') \notag \\
    &= \sum_{v_{t'}\in\{v_s,v_f\}}v_{t'}[v_s p(v_{t'},t'|v_{s},t'')p(v_{s},t'') + v_{f} p(v_{t'},t'|v_{f},t'')p(v_{f,t''})] \notag \\
    &\vdots \notag \\
    & = [v_s \eta + v_f(1-\eta)]^2 + (1-\eta)\eta(v_f-v_s)^2 e^{-(\alpha+\beta)(t'-t'')}.
\end{align}
The case $t'<t''$ is readily solved by simply making the change $t'\rightarrow t''$, $t''\rightarrow t'$. The velocity correlations can be computed, yielding
\begin{align}
    \langle v_{t'}v_{t''}\rangle = \bar{v}^2 + (1-\eta)\eta (v_f-v_s)^2 e^{-(\alpha+\beta) |t'-t''|},
\end{align}
where again $\bar{v}$ is the quenched average of the self--propulsion velocity over the velocity distribution $p(v)$. Having computed the self--propulsion velocity correlations over time, the MSD in Eq. (\ref{eq:supp msd}) can be computed. Using that $\int_{0}^{t}\int_{0}^{t}dt'dt''\;e^{-\nu |t'-t''|} = 2[(t/\nu)-(1/\nu^2)(1-e^{-\nu t})]$, the MSD writes
\begin{align}\label{eq:MSD complete}
    \langle \Delta \textbf{r}_t^2\rangle = 4Dt + 2\overline{v}^2\left(\frac{t}{D_r} -\frac{1}{D_r^2}(1-e^{-D_r t})\right) +2(1-\eta)\eta (v_f-v_s)^2\left(\frac{t}{\alpha+\beta+D_r} - \frac{1}{(\alpha+\beta + D_r)^2}(1-e^{-(\alpha+\beta+D_r)t})\right). 
\end{align}
Directly from the MSD, we find the effective diffusion coefficient $D_{\text{eff}} := \lim_{t\rightarrow \infty} \frac{\langle \Delta \textbf{r}_t ^2\rangle}{4t}$ :
\begin{align}
    D_{\text{eff}} = D + \frac{\bar{v}^2}{2D_r} + \frac{(1-\eta)\eta (v_f-v_s)^2}{2 (\alpha+\beta +D_r)}.
\end{align}
Note how, up to order $\mathcal{O}(\frac{\alpha+\beta}{D_r})$, the effective diffusion coefficient can be written as follows,
\begin{align}\label{eq:eff diff coeff big dr}
    D_{\text{eff}} = D + \frac{\eta v_s^2 + (1-\eta)v_f^2}{2D_r} + \mathcal{O}\left(\frac{\alpha+\beta}{D_r}\right) \equiv D + \frac{1}{2D_r}\overline{(v^2)} + \mathcal{O}\left(\frac{\alpha+\beta}{D_r}\right),
\end{align}
where the average $\overline{v^2}$ is done over the self--propulsion velocity distribution $p(v) = \eta \delta(v-v_s) + (1-\eta)\delta(v-v_f)$ again. %This is a quite important step into believing the effective Péclet number story later on with delay.
Indeed, Eq. (\ref{eq:eff diff coeff big dr}) can be written up to order $\mathcal{O}(\frac{\alpha+\beta}{D_r})$ in a very compact form which includes the averaged Péclet number,
\begin{align}\label{eq:eff diff coff from average peclet}
    \frac{D_{\text{eff}}}{D} = 1 + \overline{\text{Pe}} + \mathcal{O}\left(\frac{\alpha+\beta}{D_r}\right), \quad \overline{\text{Pe}} \equiv \overline{\left(\frac{v^2}{2DD_r}\right)} = \int_{v}\left(\frac{v^2}{2DD_r}\right)p(v)\;dv,
\end{align}
where again $p(v) = \eta \delta(v-v_s) + (1-\eta)\delta(v-v_f)$. This heuristically comes to say that up to order $D_r^{-1}$ the effective diffusion coefficient in units of the thermal diffusion coefficient in a mixture of particles moving at different self--propulsion velocities with a given distribution can be obtained by simply averaging the Péclet number over the sample. The scaling form of the effective diffusion coefficient in terms of the localization $\eta$ presented in the main text is readily obtained from either Eq. (\ref{eq:eff diff coeff big dr}) or Eq. (\ref{eq:eff diff coff from average peclet}),
\begin{align}\label{eq:scaling eff diff coeff}
    \frac{D_{\text{eff}}}{D}\frac{1}{1+\text{Pe}_s} = \frac{\eta}{1-\eta} + \mathcal{O}\left(\frac{\alpha+\beta}{D_r}\right).
\end{align}
In the case in which the rates $\alpha,\beta$ become comparable to $D_r^{-1}$, i.e. $\frac{\alpha+\beta}{D_r} \sim 1$, the effective diffusion coefficient might be written as follows,
\begin{align}
    D_{\text{eff}} \Big|_{\frac{\alpha+\beta}{D_r} \sim 1} \approx D_{\text{eff}}^{(0)} - \frac{D}{2} (1-\eta)\eta (\sqrt{\text{Pe}_f} - \sqrt{\text{Pe}_s})^2, \quad D_{\text{eff}}^{(0)} = D(1+\text{Pe}_s)\left(\frac{\eta}{1-\eta}\right)
\end{align}
Thus, when $\alpha,\beta$ become comparable to $D_r$ the deviation from the prediction $D_{\text{eff}}^{(0)}$ writes
\begin{align}
    \frac{D_{\text{eff}} \Big|_{\frac{\alpha+\beta}{D_r}\sim 1} - D_{\text{eff}}^{(0)}}{D} \frac{1}{(\sqrt{\text{Pe}_f} - \sqrt{\text{Pe}_s})^2} \approx - \frac{1}{2}(1-\eta)\eta.
\end{align}

\section*{III. Delay \& effective Péclet number theory}
The model presented for an ABP with sensory delay in a landscape as presented in the main text is
\begin{align}
    \begin{cases}
        \displaystyle\dot{\textbf{r}}_t = v(\textbf{r}_t, t)\hat{\nu}(\theta_t) + \sqrt{2D}\boldsymbol{\eta}(t) \\
        \displaystyle\dot{\theta}_t = \sqrt{2D_r} \chi(t)\\
        \displaystyle \dot{v}(\textbf{r}_t,t) = - \frac{1}{\tau}(v(\textbf{r}_t, t) - u(\textbf{r}_t)),
    \end{cases}
\end{align}
where $\tau$ now represents the characteristic adaptation time associated to the presence of sensory delay (or delay time). The self--propulsion velocity $v(\textbf{r}_t, t)$ now depends not only on time, but also on the trajectory of the particle and is the result of the convolution of an exponential kernel associated to the presence of sensory delay. Particularly,
\begin{align}
    v(\textbf{r}_t, t) = v_0e^{-t/\tau} + \tau^{-1}\int_{0}^{t} e^{-\frac{t-s}{\tau}}u(\textbf{r}_s)\;ds,
\end{align}
where we have assumed the particle to have an arbitrary self--propulsion velocity at $t_0 = 0$, i.e. $v(\textbf{r}_0, 0) = v_0$. When the solution to $v(\textbf{r}_t, t)$ is plugged into the Langevin equations of the model, the following is obtained,
\begin{align}
    &\dot{\textbf{r}}_t = \left(v_0e^{-t/\tau} + \tau^{-1}\int_{0}^{t} e^{-\frac{t-s}{\tau}} u(\textbf{r}_s)\;ds\right)\hat{\nu}(\theta_t) + \sqrt{2D}\boldsymbol{\eta}(t)\\
    &\dot{\theta} = \sqrt{2D_r}\chi(t),
\end{align}
which manifests the presence of what is known as \textit{distributed delay} in the translational dynamics \cite{Loos2021}. The self--propulsion of an ABP on a homogeneous landscape in which $u(\textbf{r}_t) = u_0$ is constant is simply an exponential relaxation to the target velocity $u_0$, that is $v_{t} = u_0 + (v_0-u_0)e^{-t/\tau}$. Inside a checkerboard sharp landscape, the situation is somewhat more involved. If the checkerboard consists of squares of a given size $L$ with alternating target velocities $u_{1}$ and $u_2$, the self--propulsion velocity of the particle at time $t$ can be written as,
\begin{align}\label{eq:delay self-propulsion speed time}
    v_n(t) = u_{\pi(n)} + (v_{n-1}(t_{n-1}) - u_{\pi(n)})e^{-(t-t_{n-1})/\tau}, \quad\text{with}\quad  v_k(t_k) = u_{\pi(k)}+(v_{k-1}(t_{k-1}) - u_{\pi(k)})e^{-(t_k - t_{k-1})/\tau},
\end{align}
if one assumes that starting from an initial position, $\textbf{r}_0$ at time $t = 0$, the particle changes region at times $t_1, t_2, \dots$, and so on. Here $\pi(k)$ simply denotes the target velocity index after jump $k$, i.e. if a particle sits initially with target velocity $u_1$ then $u_{\pi(1)} = u_{2}$, $u_{\pi(2)} = u_1$ and so on. Note how the residence times $T_k = t_{k}-t_{k-1}$ are random and distributed according to both the rotational and spatial dynamics but also the \textit{delay time} (making the dynamics a feedback loop). The hierarchy can be solved up to arbitrary order, so that the final solution depending on time writes
\begin{align}\label{eq:spv checkerboard with delay}
    v_{n}(t) = u_{\pi(n)} - \sum_{k=2}^{n}\Delta u_{k}e^{-\frac{t-t_{k-1}}{\tau}} + (v(t_0) - u_{\pi(1)})e^{-\frac{t-t_0}{\tau}},
\end{align}
where $v(t_0)$ is the particle's initial velocity and where $\Delta u_k = u_{\pi(k) }-u_{\pi(k-1)}$ is the discontinuity jump over the target velocities in the lands -- so that $\Delta u_k = \pm |(u_1 - u_2)|$ for any $k$. Note how $t-t_{k-1}$ can be written in terms of the residence times $T_{k}$, indeed, $t-t_{k-1} = t-t_{n-1} + T_{n-1} + T_{n-2} + \dots T_{k}$. Note how if the landscape is homogeneous, i.e $u_1 = u_2 \equiv u_0$, then $\Delta u_k = 0$ for all $k$, and $v_{n}(t) = u_0 + [v_0 - u_0]e^{-(t-t_0)/\tau}$. It is essential to understand the distribution of occupation times $T_k$ to understand the dynamic role of delay.

%It is important to note how, while once the stationary state was reached in a delay-free sharp checkerboard landscape, the distribution of self--propulsion velocities was directly related to the occupation fractions of slow and fast regions $p(v) = \eta \delta(v-v_s) + (1-\eta)\delta(v-v_f)$. 
Note that while the velocity of a particle in a delay--free sharp landscape is discrete and takes only values $v_s$ and $v_f$ (modulo Gaussian fluctuations provided by thermal noise), in the presence of finite sensory delay, this is not going to be the case anymore. Indeed the self--propulsion velocity in the stationary state when delay is present becomes a continuous random variable governed by the sequence of exponentially weighted changes in Eq. (\ref{eq:spv checkerboard with delay}). The attributed distribution of the self--propulsion velocity in the presence of delay $p_{\tau}(v)$ should recover the discrete nature of the landscape in the limit of vanishing delay, i.e. $\lim_{\tau \rightarrow 0} p_{\tau} = p(v)$ in the distributional sense.

Even though the actual shape of the distribution of self--propulsion velocities at any given delay time $\tau$ is unknown, we may consider the following; the total distribution is the result of the addition of the distribution of self--propulsion velocities in slow and fast regions with the proper normalization, i.e. $p_{\tau}(v) = c_{\tau}p_{s,\tau}(v) + (1-c_{\tau})p_{f,\tau}(v)$ where $p_{s,\tau}(v),p_{f,\tau}(v)$ are properly normalized. %\footnote{Eventhough the decomposition of a distribution as a linear combination of two different distributions is in theory not unique, we will assume uniqueness holds provided they need to comply with strong conditions because of the presence of the landscape}. 
Particularly, $c_{\tau \rightarrow 0} = \eta$ and $p_{s(f),\tau\rightarrow0}(v) = \delta(v-v_{s(f)})$ distributionally speaking. The decomposition of the distribution of self--propulsion velocities is useful in the following sense; in Eq. (\ref{eq:eff diff coff from average peclet}) the effective diffusion coefficient is written in terms of the sample averaged Péclet number $\overline{\text{Pe}}$ over the distribution $p(v)$, which in the absence of delay becomes $\overline{\text{Pe}} = \eta \text{Pe}_s + (1-\eta)\text{Pe}_f$. If the average is promoted to the average over the distribution when delay is present in the landscape, then $\overline{\text{Pe}} = c_{\tau}\overline{\text{Pe}_{s}^{\tau}} + (1-c_{\tau})\overline{\text{Pe}_{f}^{\tau}}$, where 
\begin{align}
    \overline{\text{Pe}_{s}^{\tau}} = \int_{v}\left(\frac{v^2}{2DD_r}\right)p_{s,\tau}(v)\; dv, \quad \overline{\text{Pe}_f^{\tau}} = \int_v \left(\frac{v^2}{2DD_r}\right) p_{f,\tau}(v)\;dv.
\end{align}
Thus, taking as \textit{ansatz} the validity of the Eq. (\ref{eq:eff diff coff from average peclet}) even in the presence of delay (see Sec. V where the validity of this assumption is discussed), the effective diffusion coefficient up to order $\mathcal{O}(D_r^{-1})$ becomes
\begin{align}
    \frac{D_{\text{eff}}}{D} = 1 + c_{\tau} \overline{\text{Pe}_s^{\tau}} + (1-c_{\tau}) \overline{\text{Pe}_{f}^{\tau}} + \mathcal{O}\left(D_r^{-1}\right).
\end{align}
Consequently, the scaling in Eq. (\ref{eq:scaling eff diff coeff}) can be shown to hold provided that $c_{\tau}$ is properly defined. Particularly,
\begin{align}
    \frac{D_{\text{eff}}}{D} \frac{1}{1+\overline{\text{Pe}_s^{\tau}}} = \frac{\eta_{\tau}}{1-\eta_{\tau}} + \mathcal{O}\left({D_r}^{-1}\right), \quad \text{when}\; c_{\tau} \equiv \frac{g_{\tau}}{1+g_{\tau}}, \; g_{\tau} \equiv \sqrt{\frac{1+\overline{\text{Pe}_f^{\tau}}}{1+\overline{\text{Pe}_s^{\tau}}}} \; \text{and} \; \eta_{\tau} \equiv \frac{g_{\tau}}{1+g_{\tau}}.
\end{align}
Thus, the presence of delay can be mapped into the original theory by simply promoting the Péclet numbers in the slow and fast region into the effective (average) Péclet numbers $\text{Pe}_s \rightarrow \overline{\text{Pe}_s^{\tau}}$ and $\text{Pe}_f \rightarrow \overline{\text{Pe}_{f}^{\tau}}$ so that
\begin{align}
    \frac{\eta}{1-\eta} = \sqrt{\frac{1+\text{Pe}_f}{1+\text{Pe}_s}} \equiv g &\Longrightarrow \frac{\eta_{\tau}}{1-\eta_{\tau}} = \sqrt{\frac{1+\overline{\text{Pe}_f^{\tau}}}{1+\overline{\text{Pe}_s^{\tau}}}}\equiv g_{\tau} \label{eq:eff peclet theory localization} \\
    \frac{D_{\text{eff}}}{D}\frac{1}{1+\text{Pe}_s} = \frac{\eta}{1-\eta} + \mathcal{O}\left({D_r}^{-1}\right) &\Longrightarrow \frac{D_{\text{eff}}}{D}\frac{1}{1+\overline{\text{Pe}_s^{\tau}}} = \frac{\eta_{\tau}}{1-\eta_{\tau}} + \mathcal{O}\left({D_r}^{-1}\right). \label{eq: eff peclet theory eff diff coeff}
\end{align}
Some comment on the effective Péclet numbers are provided in the next section.

\section*{IV. Justification of the telegraph approximation}\label{sec:justification telegraph}
The residence times of any simple asymmetric telegraph process are easily obtained after finding the solution to the master equation (ME) in Eqs. (\ref{eq:me f}, \ref{eq:me s}) with absorbing boundary conditions. The residence time distributions are $\psi_{f}(t) = \beta e^{-\beta t}$ and $\psi_{s}(t) = \alpha e^{-\alpha t}$, meaning the residence times of the slow and fast states are exponentially distributed, $T_{f} \sim \text{exp}(\beta)$, $T_{s} \sim \text{exp}(\alpha)$, so that $\langle T_{f} \rangle = \beta^{-1}$, $\langle T_{s}\rangle = \alpha^{-1}$.\\

While the choice to model swimmers in a sharp motility landscape might be intuitively reasonable because of the underlying dynamics and the resulting switching of the self--propulsion speed, it is not a formal step. However, and as shown in Fig. \ref{fig:residence time distributions} (a), the residence times in the stationary state of a mixture of non--interacting particles in a checkerboard landscape seem to be exponentially distributed. This is further shown in panels b) and c) of Fig. \ref{fig:residence time distributions}. While the scaling is not exactly precise when using the average residence in order to universally scale all the distributions as shown in Fig. \ref{fig:residence time distributions} (b), the collapse works when fitting the tails to an exponential distribution with parameter $\tilde{T}_{\tau}$. This is the case because the average residence time is only a good estimate of the parameter of the exponential when the full distribution is exponential, and in this case only the tails seem to be distributed exponentially.

\begin{figure}[h!]
    \centering
    \includegraphics[width=0.9\linewidth]{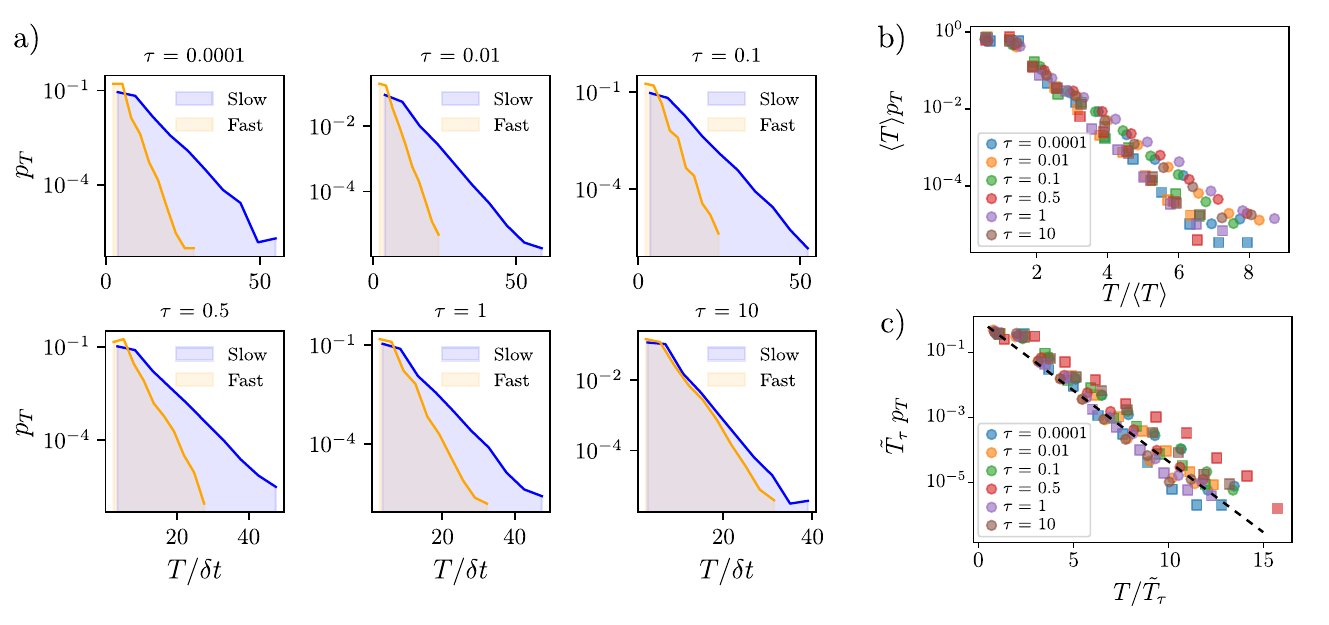}
    \vspace{-0.6cm}
    \caption{\textbf{Residence time distributions.} (a) Residence time distribution of a checkerboard landscape with domain size $\ell = \frac{1}{4}$ in a simulation box of total length $L = 1$ and target velocities $v_f = \frac{1}{4}$, $v_s = \frac{1}{8}$, $\tau_r = 1$ and $\tau_D = 10^4$ in the presence of time delay of magnitudes $\tau = 10^{-4}, 10^{-2}, 10^{-1}, 0.5, 1, 10$. The time is measured in units of the timestep of the simulation. For any $\tau$, the distribution of the occupation times $p_T$ seems to be linear in lin--log scale, signaling that the tails are exponentially distributed. (b) Collapse of the distributions for all $\tau$ in the fast (with target velocity $v_f$) and slow (with target velocity $v_s$) using the average value of the residence $\langle T \rangle$. Discrepancies from the real exponential distribution comes because the full distribution of residence time is not an exponential, only the tails are. The squares represent data corresponding to the residence times of fast regions and the circles to the residence of slow ones. (c) \textit{Ídem}, collapse of the distributions by fitting the exponential parameter at the tails of the distributions, assuming $T \sim \text{Exp}(\tilde{T}_{\tau})$, so that at the tails $\tilde{T}_{\tau}p_T \sim e^{-T/\tilde{T}_{\tau}}$. The dashed line corresponds to an exponential distribution with unit mean.}
    \label{fig:residence time distributions}
\end{figure}
\vspace{-0.5cm}

\section*{V. Delay \& effective Péclet numbers: validity, properties \& relevance}\label{sec:v}
The idea behind using a phenomenological description with effective Péclet numbers is supported by the presence of the averages over the self--propulsion speed distribution $p(v)$ in the MSD, Eq. (\ref{eq:MSD complete}), and the effective diffusion coefficient, Eq. (\ref{eq:eff diff coff from average peclet}). Now, since $D_{\text{eff}}/D \approx 1 + \overline{\text{Pe}}$ is limited to order $D_r^{-1}$, the extension of this result to the presence of delay will only be valid at up to order $D_r^{-1}$. Nothing can be said further beyond, because the dynamics of the self--propulsion speed qualitatively differ from that of a telegraph process, in the presence of delay.

\subsection*{A. Regime of applicability and extension in the presence of delay}
The shape of the effective diffusion coefficient is the result of an average of the Péclet number in the limit in which $\frac{\alpha+\beta}{D_r} \ll 1$. Note how since $\alpha = \langle T_s\rangle^{-1}$ and $\beta = \langle T_f\rangle^{-1}$ as discussed in Sec. \ref{sec:justification telegraph}, this requires that $\tau_r [\langle T_s\rangle^{-1} + \langle T_f\rangle^{-1}] \ll 1$, where $\langle T_s \rangle$ and $\langle T_f\rangle$ are the expected residence times of slow and fast states respectively. Note how, for any finite $D_r$ the expected residence times will always be greater than the ballistic characteristic times, i.e. $\langle T_{s}\rangle \geq \tau_{\ell,s}$ and $\langle T_f \rangle \geq \tau_{\ell,f}$. Hence, the theory until order $D_{r}^{-1}$ will be valid if
\begin{align}
    \tau_r \left(\frac{1}{\langle T_s\rangle } + \frac{1}{\langle T_f \rangle}\right) \leq \tau_r \left( \frac{1}{\tau_{\ell,s}} + \frac{1}{\tau_{\ell,f}}\right) \ll 1.
\end{align}
For any landscape with fixed velocities $v_s$, $v_f$, domain length $\ell$ (that is, with fixed $\tau_{\ell,s}, \tau_{\ell,f}$) and finite $D_r$ and $D$, $D_{\text{eff}}/D \approx 1 +  \overline{\text{Pe}}$ will hold if we consider only
\begin{align}
    \tau_r \leq\tau_r^{(\varepsilon)} = \varepsilon \frac{\tau_{\ell,s} \tau_{\ell,f}}{\tau_{\ell,s} + \tau_{\ell,f}},
\end{align}
where $\varepsilon \ll 1$. From the original theory without delay one may show that $\eta$ increases with increasing $\tau_r$, meaning the values of $\eta$ for which the zero order approximation $\frac{D_{\text{eff}}}{D} \approx 1+ \overline{\text{Pe}}$ holds are the ones for which
\begin{align}
    \eta \leq \eta^{(\varepsilon)} = \frac{g^{(\varepsilon)}}{1+g^{(\varepsilon)}}, \quad g^{(\varepsilon)} = \sqrt{\frac{1 + \frac{\varepsilon}{2}\frac{\tau_{\ell,s}}{\tau_{\ell,f}} \frac{\tau_D}{\tau_{\ell,s} + \tau_{\ell,f}}}{1+ \frac{\varepsilon}{2}\frac{\tau_{\ell,f}}{\tau_{\ell,s}} \frac{\tau_D}{\tau_{\ell,s}+\tau_{\ell,f}}}},
\end{align}
where again $\varepsilon\ll 1$. In a practical context, one may take $\varepsilon \sim 10^{-2}$. The extension of the theory in the presence of finite sensory delay is expected to provide a good description only for $\eta$ smaller than $\eta^{(\varepsilon)}$. Note how in the case in which the ballistic time--scales $\tau_{\ell,f}$ and $\tau_{\ell, s}$ are of the same order, i.e. $\tau_{\ell,s}\sim \tau_{\ell,f} \sim \tau_{\ell}$, then the zero--order approximation $D_{\text{eff}}/D \approx 1 + \overline{\text{Pe}}$ is valid when $\tau_r \lesssim \varepsilon \tau_{\ell}$, meaning $\tau_r/\tau_{\ell} \ll 1$ or $\tau_{\ell}/\tau_r \gg 1$.

\subsection*{B. Properties of delay induced effective Péclets}
The decomposition of the self--propulsion speed $p_{\tau}(v) = c_\tau p_{s,\tau} + (1-c_{\tau})p_{f,\tau}$ comes directly from the idea that one may consider the self--propulsion speed distribution on fast and slow regions separately.  When time delay is not relevant, i.e. when the time--scales involved in the landscape are greater than $\tau$ (see Tab. (\ref{tab:relevance of delay table})), the self--propulsion velocity almost instantaneously adapts to changes induced by spatial modulations, so that the self--propulsion speeds do not change significantly. This effectively means that $p_{\nu,\tau} \approx \delta(v-v_{\nu})$ with $\nu = s,f$ when $\tau \approx 0$. As a consequence, the effective Péclet numbers are simply $\overline{\text{Pe}_{\nu}^{\tau}} \approx \text{Pe}_{\nu}$ for $\nu = s,f$. When time delay $\tau$ becomes comparable to the other time--scales of the system, a particle may enter and leave a given domain without having fully adapted to the self--propulsion speed of that given region. Since the particle is going fast before entering a slow region, the propulsion velocity at any time before exit is greater than $v_s$. The same thing happens when a particle enters a fast region, since at the beginning it is moving at a lower speed, the propulsion velocity at any time before exit will be lower than $v_f$. This comes to say that, as soon as $\tau$ becomes comparable to the \textit{residence} times of fast and slow regions, the propulsion speed does not fully adapt to either $v_s$ or $v_f$. This is depicted with an example in Fig. \ref{fig:effects of delay}, were both the stationary density patterns and velocity distributions are shown for different values of $\tau$ (with $\tau$ increasing to the right, each column has the same value of $\tau$). In the limiting case in which delay is greater than any scale involved in the landscape, the particles barely see the spatial modulations and as a result the system behaves effectively homogeneous. It can be shown that in this case, if the landscape is symmetric, the self--propulsion speed effectively tends to the arithmetic mean of the velocities of the landscape, $u = \frac{v_s + v_f}{2}$. This can be shown, on average, as follows; directly from Eq. (\ref{eq:delay self-propulsion speed time}) it follows that
\begin{align}
    \mathbb{E}[v_n(t_n)] \equiv \mathbb{E}[v_n] = u_{\pi(n)}(1-\mathbb{E}[e^{-T_n/\tau}]) + \mathbb{E}[v_{n-1}]\mathbb{E}[e^{-T_n/\tau}],
\end{align}
where we have used the independence of the value $v_{n-1}$ with the residence time $T_n = t_{n}-t_{n-1}$. Note how, in the large time delay limit $\tau \gg 1$ the distribution of the time residences of slow and fast regions converge, making the distribution of the residence of slow and fast regions identical, see Fig. \ref{fig:effects of delay} (a) ($\tau = 10$ panel). This means that the residence times are sampled exponentially and identically $T_{k} \sim \text{exp}(\tau_{\infty}^{-1})$. Taking this into account, 
\begin{align}
    \mathbb{E}[v_n] = u_{\pi(n)}\left(1 - \frac{\tau}{\tau_{\infty} + \tau}\right) + \frac{\tau}{\tau_{\infty} + \tau} \mathbb{E}[v_{n-1}].
\end{align}
Defining now $z = \frac{\tau}{\tau_{\infty} + \tau}$ and iterating, one finds
\begin{align}
    \mathbb{E}[v_n] = (1-z) \sum_{k=0}^{n-1}u_{\pi(n-k)} z^{k} + z^{n}\mathbb{E}[v_0].
\end{align}
Furthermore, assuming that the process is stationary and that the average velocities are the same, since the residence times are identically distributed, so that $\mathbb{E}[v_n] = \mathbb{E}[v_0]$, one may find a closed expression for the average velocity. Indeed, plugging $\mathbb{E}[v_n] = \mathbb{E}[v_0] \equiv \mathbb{E}[v]$ takes to
\begin{align}
    \mathbb{E}[v] = \frac{1-z}{1-z^n} \sum_{k = 0}^{n-1}u_{\pi(n-k)} z^{k}.
\end{align}
Since the process has been assumed to be stationary one may take the $n\rightarrow \infty$ limit of the sum after choosing a particular sequence, i.e. $u_{\pi(1)} = u_1, u_{\pi(2)} = u_2, u_{\pi(3)} = u_1, \dots$ and so on. After doing that, one has
\begin{align}
     \mathbb{E}[v] \equiv \lim_{n\rightarrow \infty}\mathbb{E}[v] = (1-z) \left(u_1 \sum_{k=0}^{\infty} z^{2k+1} + u_2 \sum_{k=1}^{\infty}z^{2k}\right) = \frac{z(u_1 + z u_2)}{1+z},
\end{align}
since $|z| < 1$ and where we have just simply separated the sum into even and odd integer numbers according to the chosen sequence $u_{\pi(1)} = u_1, u_{\pi(2)} = u_2, \dots$ and so on. Note how the residence times are only identically distributed for big enough delay times, meaning this is only valid when $\tau \gg 1$. Taking the limit in which $\tau/\tau_{\infty} \rightarrow \infty$ or $z \rightarrow 1$ it directly follows that
\begin{align}
    \mathbb{E}[v]_{\tau \gg 1} \equiv \lim_{z \rightarrow 1} \mathbb{E}[v] = \lim_{z\rightarrow 1} \frac{z(u_1 + zu_2)}{1+z} = \frac{u_1 + u_2}{2}.
\end{align}
As a consequence of this and even though we don't explicitly know the shape of the effective Péclets $\overline{\text{Pe}_{s,f}^{\tau}}$, they comply with the following,
\begin{align}
    \lim_{\tau \rightarrow 0} \overline{\text{Pe}_{s,f}^{\tau}} = \text{Pe}_{s,f}, \quad \lim_{\tau\rightarrow \infty} \overline{\text{Pe}_{s,f}^{\tau}} = \left(\frac{\sqrt{\text{Pe}_s} + \sqrt{\text{Pe}_f}}{2}\right)^2 \equiv \text{Pe}_{\infty},
\end{align}
This is consistent with the claim that for very small sensory delay or time delay $\tau\approx 0$, $p_{\tau}(v) \sim \eta\delta(v-v_s) + (1-\eta)\delta(v-v_f)$ while for big enough time delays, $\tau \gg 1$, $p_{\tau} \sim \delta(v-\frac{v_s+v_f}{2})$. This is depicted in the lower panels of Fig. \ref{fig:effects of delay} where histograms of the instantaneous velocities are shown in the stationary state for different values of $\tau= 0.1, 0.5, 1, 10$. It is clear how for small delay the distribution is peaked around the target velocities provided by the landscape, and how the distribution is reshaped until it becomes unimodal and peaked around $\frac{v_s+v_f}{2}$ for large enough $\tau$.

\begin{figure}[h!]
    \centering
    \includegraphics[width=0.925\linewidth]{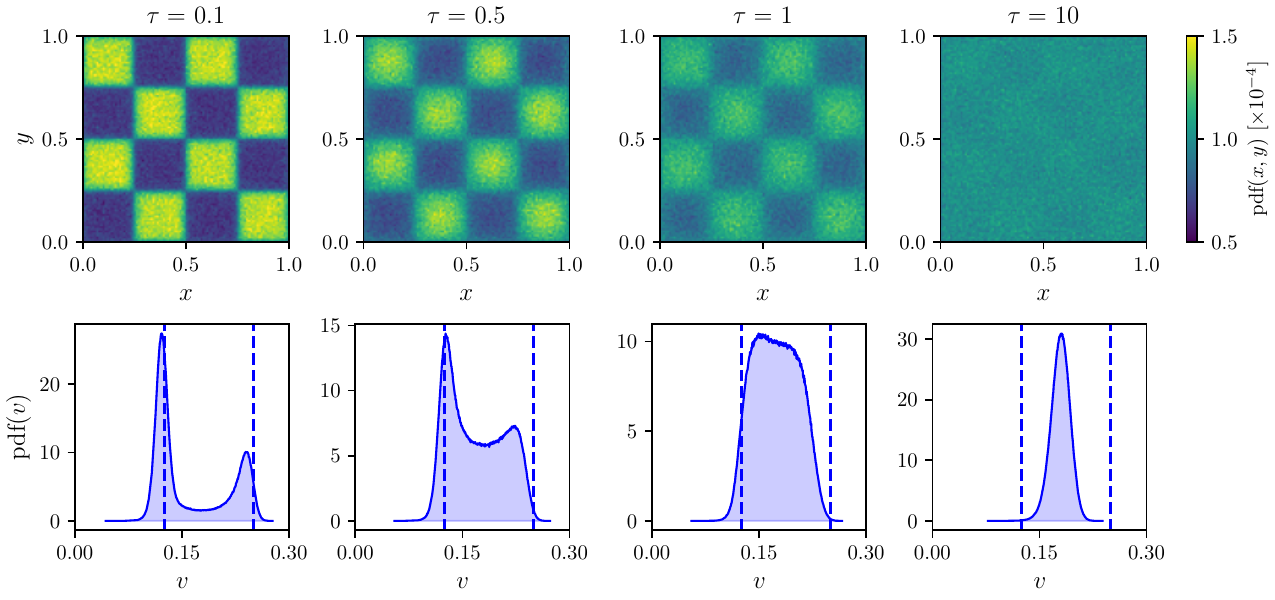}
    \vspace{-0.4cm}
    \caption{\textbf{Sensory delay, reshaping of stationary density patterns and instantaneous velocity distributions.} Stationary density patterns in a checkerboard landscape of domain size $\ell = \frac{1}{4}$ inside a box of total length $L=1$ with self--propulsion velocities $v_s = 0.125$, $v_f = 0.25 $ and $D_r = 1$, $D = \frac{1}{4}\times 10^{-4}$ (top row) and the corresponding instantaneous velocity distribution (bottom row) for different values of $\tau = 0.1, 0.5, 1, 10$.}
    \label{fig:effects of delay}
\end{figure}

The modulation of stationary density patterns through sensory delay presents interesting behavior. In \cite{Tpfer2025} it was shown how the occupation fraction $\eta$, or analogously the \textit{localization} $\frac{\eta}{1-\eta}$ may or may not be optimized at a finite value of delay $\tau^*$ depending on statistic properties, such as the self--propulsion velocities in slow and fast regions, i.e. $v_s$ and $v_f$, and the size of the checkerboard domains, $\ell$. This is shown in Fig \ref{fig:non-monotonic behaviour} (a). Similar behavior is seen for the effective diffusion coefficient, which may be optimized at a finite value of sensory delay $\tau^*$ depending on the landscape's stationary properties. Interestingly, the set that optimizes the localization $\frac{\eta}{1-\eta}$, i.e. $\ell = 40, 60, 80, 100$, lies close to the frontier $\tau_r \sim \tau_{\ell}$ as shown in Fig \ref{fig:non-monotonic behaviour} (c), where the landscape changes from a ballistic dominant regime to a diffusion dominant one. On the other hand, the set that does not optimize the localization, i.e. $\ell = 5, 10$, optimizes the effective diffusion coefficient at a finite value of $\tau$. This set lies close to the frontier $\tau \sim \tau_r$, when $\tau_{\ell}/\tau_{r}\ll 1$ (ballistic region).

\begin{figure}[h!]
    \centering
    \includegraphics[width=0.95\linewidth]{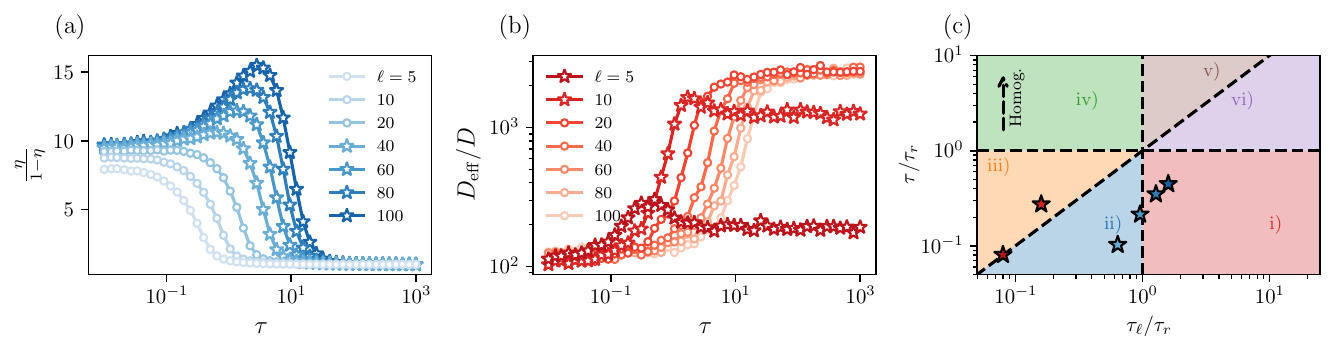}
    \vspace{-0.4cm}
    \caption{\textbf{Non--monotonic behaviour and regime map.} (a) Localization $\frac{\eta}{1-\eta}$ as a function of sensory delay $\tau$ for a landscape of fixed velocities $v_s = 1$, $v_f = 10$, $D_r = 6.26$ and $D = 1.25\times 10^{-1}$ of a system of total size $L = 10^3$ with checkerboard domains of size $\ell = 5,10,20,40,60,80,100$. The circles represents data that is not optimized at finite $\tau$ and the stars data for which the localization is indeed optimized at a finite value of sensory delay $\tau^*$. The data is extracted from \cite{Tpfer2025} and serves here as example to prove a point. (b) \textit{Ídem} but now the effective diffusion coefficient for the same landscape as a function of the sensory delay $\tau$ is shown. Notice how the set that in panel (a) optimizing the localization $\frac{\eta}{1-\eta}$ now does not optimize the effective diffusion, while the set that didn't optimize the localization in (a) optimizes the effective diffusion coefficient. (c) Map in which we show where in the time--scale separation table Tab. (\ref{tab:relevance of delay table}) lie the optimal values in panels (a) and (b). Note how only the time--scales of fast domains are shown, the slow regions' time--scales always lay at the right of the present scatters.}
    \label{fig:non-monotonic behaviour}
\end{figure}

\subsection*{C. Time--scale separation and relevance of delay}
A table is left here where the separation of scales, the relevance of sensory delay and the applicability of the extension of the theory in the presence of delay is summarized (translation thermal diffusion is considered to be the largest time--scale, $\tau_{\ell}, \tau_{r}, \tau \ll \tau_{D}$).

\begin{table}[h!]
    \centering
    \begin{tabular}{||ccccc||}
    \hline
       Scale sep. & Regime & Theory ($\tau = 0$) & Delayed extension & Properties\\
    \hline\hline
       (i) $\tau<\tau_r<\tau_{\ell}$ & Diffusive ($\delta < 1$, delay negligible) & $\newcheckmark$ & $\newcheckmark$ & Optimal $\eta$ at finite $\tau^*$\\
       (ii) $\tau<\tau_{\ell}<\tau_{r}$ & Ballistic ($\delta >1$, delay negligible) & $\newcheckmark$ & $\newcrossmark$ & Optimal $\eta$ at finite $\tau^*$\\
       (iii) $\tau_{\ell}<\tau<\tau_{r}$ & Ballistic ($\delta >1$, delay relevant) & $\newcrossmark$ & $\newcrossmark$ & Optimal $D_{\text{eff}}$ at finite $\tau^*$\\
       (iv) $\tau_{\ell}<\tau_{r}<\tau$ & Ballistic ($\delta > 1$, homogeneous through $\tau$) & $\newcrossmark$ & $\newcrossmark$ & (Not studied) \\
       (v) $\tau_r<\tau_{\ell}<\tau$ & Diffusive ($\delta < 1$, homogeneous through $\tau$) & $\newcrossmark$ & $\sim \newcheckmark$ & (Not studied) \\
       (vi) $\tau_r<\tau<\tau_{\ell}$ & Diffusive ($\delta < 1$, homogeneous through $\tau$) & $\newcrossmark$ & $\sim \newcheckmark$ & (Not studied)\\
    % \hline\hline
    %   $\tau_{\ell}< \tau_{r}$ & Optimal $\eta_{\tau}$ at $\tau^* = 0$ \\
    % \hline\hline
    \hline   
    \end{tabular}
    \caption{\textbf{Time--scale separation, regimes and applicability.} Table showcasing all the possible regimes through a time--scales separation and summarizing phenomenology and applicability of the extension to the theory in the presence of delay. It is assumed that the condition relating the bulk density ratios is satisfied, meaning the domain sizes are big enough so that $\lambda/\ell \ll 1$, with $\lambda = \sqrt{D/D_r}/\sqrt{1+\text{Pe}}$. Here $\delta_{s,f} \sim \frac{\tau_r}{\langle T_{s,f}\rangle}$.}
    \label{tab:relevance of delay table}
\end{table}

Note how ``not studied" in Tab. (\ref{tab:relevance of delay table}) means no simulations or experiments are performed in this cases because the simulations are experiment--based and the experimental set up does not reach regimes (iv), (v) and (vi) for now. The use of $\sim \newcheckmark$ means we hypothesize the theory to work in those regimes even though the theory has not been tested in these regimes.

\subsection*{D. Measurement of effective Péclet numbers in the presence of delay}
Extracting the effective Péclet numbers induced by the presence of delay through the decomposition of the self--propulsion velocity in slow and fast regions and computing the averages is not practical. Indeed, consider an ensemble of ABPs that self--propel with a randomly distributed self--propulsion velocity $v_0$, which has distribution $p(v_0)$. The average over the disorder in $v_0$ of the MSD can be written as,
\begin{align}
    \overline{\text{MSD}}(t) = \int dv_0\;\text{MSD}_{v_0}(t)\;p(v_0) = 4Dt + \frac{2 \overline{v_0^2}}{D_r^2}(D_rt - 1  + e^{-D_r t}),
\end{align}
so that at small times
\begin{align}
    \overline{\text{MSD}}(t) = 4Dt + \overline{v_0^2}t^2 + \mathcal{O}(t^3) \equiv 4Dt + 2D_r \overline{\text{Pe}} \;t^2 + \mathcal{O}(t^3).
\end{align}
As a consequence, in the presence of delay, one may extract the effective Péclet numbers in slow and fast regions by simply tracking the MSD in slow and fast regions respectively and fitting a parabola, which has coefficients,
\begin{align}
    \overline{\text{MSD}}_{s,f}(t) = 4Dt + 2D_r \overline{\text{Pe}_{s,f}^{\tau}}\;t^2 + \mathcal{O}(t^3).
\end{align}

\end{widetext}

\end{document}